\documentclass[12pt]{article}
\usepackage{graphicx}
\usepackage{amsmath,latexsym,epsfig,epsf,rotate}

\begin{document}

\title{Final--state radiation in electron--positron annihilation into a  pion pair }
\author{S.~Dubinsky$^{a)}$, A.~Korchin$^{b)\footnote{E-mail: korchin@kipt.kharkov.ua}}$, N.~Merenkov$^{b)\footnote{E-mail: merenkov@kipt.kharkov.ua}}$, G.~Pancheri$^{c)\footnote{E-mail: Giulia.Pancheri@lnf.infn.it}}$%
, O.~Shekhovtsova$^{b)\footnote{E-mail: shekhovtsova@kipt.kharkov.ua}}$ \\
\emph{$^{a)}$Kharkov National University, Kharkov 61077, Ukraine} \\
\emph{$^{b)}$NSC ``Kharkov Institute for Physics and Technology'',} \\
\emph{Institute for Theoretical Physics, Kharkov 61108, Ukraine } \\
\emph{$^{c)}$INFN Laboratori Nazionale di Frascati, Italy}}
\date{}
\maketitle

\begin{abstract}
The process of $e^+ e^-$ annihilation into a  $\pi^+ \pi^-$ pair with radiation
of a  photon is considered. The amplitude of the reaction $e^+ e^- \to
\pi^+ \pi^- \gamma$ consists of the model independent initial-state
radiation (ISR) and model dependent final-state radiation (FSR). The general
structure of the FSR tensor is constructed from Lorentz covariance, gauge
invariance and discrete symmetries in terms of the three invariant
functions. To calculate these functions we apply Chiral Perturbation Theory
(ChPT) with vector and axial-vector mesons.
 The contribution of $e^+ e^- \to \pi^+ \pi^- \gamma$ process to
 the muon anomalous magnetic moment is evaluated, and results are
 compared with the dominant contribution in the  framework of a hybrid model,
 consisting of VMD and point-like scalar electrodynamics.
 The developed approach allows us
 also to calculate the $\pi^+ \pi^-$ charge asymmetry.
\end{abstract}

\vspace{0.5cm}

\vspace{0.2cm} PACS: 12.20.-m; 12.39.Fe; 13.40.-f; 13.66.Bc

\vspace{0.2cm}

\section{Introduction}

\label{sec:introduction}

\hspace{0.5cm}The cross section of electron-positron annihilation into hadrons, $%
e^+ e^- \to \ hadrons$, is crucial for evaluation of the hadronic
contribution to anomalous magnetic moment (AMM) of the muon
$a_\mu^{had}$, and is at present one of the main sources of
theoretical uncertainty in the value of AMM \cite{Eidelm_1}. In
order to resolve the existing deviation of the experimental and
Standard Model prediction values of  AMM, the corresponding
hadronic contribution is needed with very high accuracy,
better than 1\%. This is especially important in view of 
a new E969 experiment, which is expected to measure AMM about three times more accurately 
the updated $%
E821$ experiment, in which the data are expected to have
AMM about three times more accurate  than now \cite{Upgm}.

The hadronic contribution to AMM cannot be reliably calculated in
the framework of perturbative QCD, because  low-energy region
dominates, and one usually resorts to dispersion relations, where
the experimental total cross section is the input. Experimentally,
the energy region from threshold to the collider beam energy is
explored at the $\Phi$-factory DA$\Phi$NE (Frascati)
\cite{Cataldi_99} and $B$-factories BaBar (SLAC) and Belle (KEK)
\cite{Sol,Benayoun_99} using the method of radiative return \cite
{Chen_75,Rr1,Rr2}. In spite of the loss in the luminosity, this
method potentially may have  advantages in systematics over the
more traditional direct scanning measurements performed at
different CM energies, such as experiments on VEPP--2M
(Novosibirsk) \cite {Akhmetshin_CMD-2} and BES (Bejing)
\cite{pekin}.

The  radiative return method  relies on a factorization of the $e^+
e^- \to \ hadrons + \gamma$ cross section in the product of the
hadronic cross section $\sigma(e^+ e^- \to \ hadrons)$ taken at a
reduced CM energy and a model--independent radiation function known
from Quantum Electrodynamics (QED) \cite{Rr2,Baier_65,Khoze_02}.
This factorization is valid only for  photon radiation from the
initial leptons (initial--state radiation (ISR)). The additional
contribution from  photon radiation off the final hadrons
(final-state radiation (FSR)) is model dependent and becomes a
background in the radiative return scanning measurements. That is why
the problem of the separation of ISR and FSR has become quite
important.

Different methods have been suggested to separate ISR and FSR contributions
for the dominant hadronic channel at low energies, mainly the pion pair
production process
\begin{equation}  \label{eq:e+e-}
e^-(p_1)+ e^+(p_2) \to \pi^+(p_+) + \pi^-(p_-) + \gamma(k).
\end{equation}
One of them is to choose a kinematic set up, where the photon is radiated
along the momenta of the leptons
(DA$\Phi$NE setup, \cite{Cataldi_99,Rr2} and references therein).
In these conditions the FSR contribution is suppressed.  If the
FSR background can be reliably calculated in some theoretical
model, then it can be subtracted from the experimental cross section of
process (\ref{eq:e+e-}) or incorporated in the Monte Carlo
event generator used in the analysis. Finally, the theoretical
predictions for FSR can be tested by studying the $C$--odd
interference of ISR and FSR \cite{Cataldi_99,Rr2}.


Another, and even more important reason why one should know the
FSR cross section is the fact that the next-to-leading order
hadronic contribution $ a_\mu^{had, \gamma}$ to AMM, where
an additional photon is attached to hadrons, is of the order of the
present experimental accuracy.

The FSR cross section in  process (\ref{eq:e+e-}) has been
calculated \cite{Baier_65, Czyz_03} in the  framework of  scalar QED
(sQED), in which the pions are treated as point-like particles,
and the resulting amplitude is multiplied by the pion
electromagnetic form factor $F_\pi (s)$ evaluated in the Vector Meson
Dominance (VMD) model ($s$ is the invariant $e^+e^-$ energy
squared) to account for the pion structure. In this model the
contribution of the channel $\pi^+\pi^-\gamma$ to AMM is estimated
as  $ a_\mu^{had, \gamma}= 4.3 \times 10^{-10}$ \cite{Czyz_03}.
Although sQED in some cases works well \cite{Cataldi_99,Czyz_03},
it is clear that sQED is a simplified model of a complicated
process, which may include excitation of resonances, loop
contributions, etc. In view of the above mentioned requirements
for the accuracy of theoretical predictions, further studies of
the FSR contribution are necessary.

In this paper we consider the  $e^+e^- \to \pi^+ \pi^- \gamma$ reaction
in detail, focusing on FSR. Firstly, we specify the
model-independent structure of the FSR amplitude, based on Lorentz
covariance, gauge invariance and discrete symmetries. Taking this
structure into account we rewrite the FSR contribution, as well as
the interference of ISR and FSR, in terms of the three scalar
functions $f_i$ depending on three kinematical invariants.
Secondly, the model-dependent functions $f_i$ are obtained in
the framework of Chiral Perturbation Theory (ChPT) with vector $\rho
(770)$ and axial-vector $a_1 (1260)$ mesons \cite{Ecker_89}. In
this way $f_i$ are expressed through the several constants
entering the ChPT Lagrangian.

For experimental conditions, in which the cross section integrated
over the full phase space of the two pions is required, this
integration is carried out analytically. We obtain a general
result for the cross section $ \mathrm{d}\sigma /\mathrm{d} q^2
\mathrm{d} \cos \theta $ in terms of the two scalar functions $
h_{1,2}(q^2,s)$ ($q^2$ is the invariant mass of the $\pi^+ \pi^-$
pair, $\theta$ is the angle between photon and electron momenta).

We further study the interference of  ISR and FSR  by calculating
the $\pi^+ \pi^-$ charge asymmetry. Finally, the contribution for
$\pi^+\pi^-\gamma$ channel to $ a_\mu^{had, \gamma}$ is evaluated
and results are compared with sQED predictions.

The paper is organized as follows. In Sect.~\ref{sec:formalism}
the general form of the amplitude is considered. The squared and
averaged amplitudes for ISR and FSR, as well as the interference
part are analytically calculated and the structure of the cross
section $\mathrm{d}\sigma /\mathrm{d} q^2 \mathrm{d} \cos \theta $
is studied. The invariant functions $f_i$ in framework of ChPT are
derived in Sect.~\ref{sec:final-state-ChPT}. Results of
calculations and discussion are presented in
Sect.~\ref{sec:results}. In Sect.~\ref{sec:conclusions} we draw
conclusions. In Appendix~A we discuss symmetries of the FSR tensor
and its model-independent structure. Appendix~B contains the ChPT
Lagrangian and explicit expression for the FSR tensor. In
Appendix~C the Feynman rules needed for evaluation of the
$\gamma^* \to \gamma \pi^+ \pi^-$ amplitude are specified.
Appendix~D collects expressions for scalar functions
$h_{1,2}(q^2,s)$ in ChPT. Some subtle aspects of the kinematics at
low values of $q^2$ are presented in Appendix~E. In Appendix~F the
contribution to the FSR tensor and charge asymmetry from the
''anomalous'' $\rho \to \pi \gamma$ process is calculated.

\section{Formalism for $e^{+}e^{-} \rightarrow \pi ^{+}\pi ^{-}\gamma $
reaction}

\label{sec:formalism}
\begin{figure}[tbp]
\label{fig1}
\begin{center}
\epsfig{file=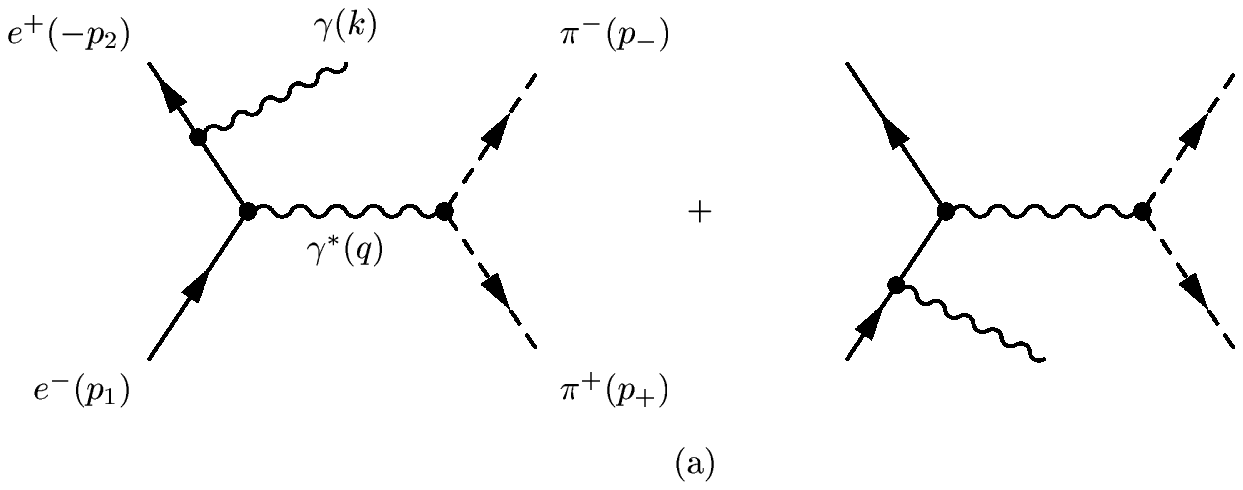,width=8cm,height=4cm} 
\epsfig{file=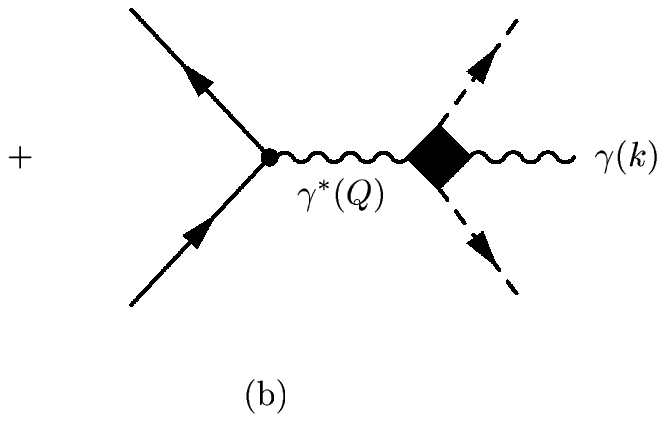,width=3.5cm,height=3.2cm}
\end{center}
\caption{Diagrams describing $e^+e^-\to\pi^+\pi^- \gamma$ process
}
\par
\end{figure}

\hspace{0.5cm} Reaction (1) is described by diagrams depicted
in Fig.~1. To analyze it we introduce 4-momenta $Q=p_{1}+p_{2}$,
$q=p_{+}+p_{-}$ and $l=p_{+}-p_{-}$. The amplitude of the reaction
depends on five independent Lorentz scalars, which can be chosen
as follows:
\begin{eqnarray}
s &\equiv &Q^{2}=2p_{1}\cdot p_{2},\ t_{1}\equiv
(p_{1}-k)^{2}-m^2_e=-2p_{1}\cdot k,\ t_{2}\equiv
(p_{2}-k)^{2}-m^2_e=-2p_{2}\cdot k,  \nonumber \\
u_{1} &\equiv &l\cdot p_{1},\ u_{2}\equiv l\cdot p_{2},  \label{eq:scalars}
\end{eqnarray}
where we neglected the  electron mass ($m_e$) in the expression for
$s$. For further reference note that other invariants are related
to those in Eqs.(\ref {eq:scalars}), for instance,
$q^{2}=s+t_{1}+t_{2}$, \ $l^{2}=4m_{\pi
}^{2}-s-t_{1}-t_{2}$, \ $q\cdot l=0$, \ $k\cdot Q=k\cdot q=-\frac{1 }{2}%
(t_{1}+t_{2})$, \ $k\cdot l=Q\cdot l=u_{1}+u_{2}$.

The amplitude of  process (\ref{eq:e+e-}) is the sum $M=M_{ISR}+M_{FSR}$,
where $M_{ISR}$ ($M_{FSR}$) describes ISR (FSR) process
\begin{equation}
M_{ISR}=\frac{e}{q^{2}}L^{\mu \nu }\epsilon _{\nu }^{\ast }l_{\mu
}F_{\pi }(q^{2}),\;\ \;\;\;\;M_{FSR}=\frac{e^{2}}{s}J_{\mu
}M_{F}^{\mu \nu }\epsilon _{\nu }^{\ast },
\label{eq:total_amplitude}
\end{equation}
where the lepton currents are given by
\begin{eqnarray}
L^{\mu \nu } &=&e^{2}\bar{u}_{s_{2}}(-p_{2})[\gamma ^{\nu }\frac{(-p%
\hspace{-0.5em}/_{2}+k\hspace{-0.5em}/+m_{e})}{t_{2}}\gamma ^{\mu }+\gamma
^{\mu }\frac{(p\hspace{-0.5em}/_{1}-k\hspace{-0.5em}/+m_{e})}{t_{1}}\gamma
^{\nu }]u_{s_{1}}(p_{1}),  \label{eq:lepton_cur_1} \\
J_{\mu } &=&e\bar{u}_{s_{2}}(-p_{2})\gamma _{\mu
}u_{s_{1}}(p_{1}), \label{eq:lepton_cur_2}
\end{eqnarray}
$F_{\pi }(q^{2})$ is the pion electromagnetic (EM) form factor, $%
\epsilon_\nu^{\ast}$ is the photon polarization vector and the tensor $%
M_{F}^{\mu \nu }$ describes the photon radiation from the final
state. This tensor is considered in detail in
Sect.~\ref{sec:final-state-ChPT} and
Appendices~A and B. In Eqs.~(\ref{eq:lepton_cur_1}) and (\ref{eq:lepton_cur_2}%
) $u_{s_{1}}(p_{1})$ and $\bar{u}_{s_{2}}(-p_{2})$ are the electron and
positron spinors with normalization: \ $\bar{u}_{s^{\prime }}(p)u_{s}(p)=-%
\bar{u}_{s^{\prime }}(-p)u_{s}(-p)=2m_{e}\delta _{ss^{\prime }}$.

The invariant amplitude squared, averaged over initial lepton polarizations
and summed over the photon polarizations\footnote{%
We use $\sum_{polar.}\epsilon _{\rho }^{\ast }\epsilon _{\sigma }=-g_{\rho
\sigma }$} can be written as
\begin{equation}
\overline{|M|^{2}}=\overline{|M_{ISR}|^{2}}+\overline{|M_{FSR}|^{2}}+2\mathrm{Re}(%
\overline{M_{ISR}M_{FSR}^{\ast }}).  \label{sq_ampl}
\end{equation}
The expressions for $\overline{|M_{ISR}|^{2}}$,
$\overline{|M_{FSR}|^{2}}$ and the interference part $\mathrm{Re}(
\overline{M_{ISR}M_{FSR}^{\ast }})$ are given in
Sects.~\ref{subsec:initial-state}, \ref{subsec:final-state} and
\ref {subsec:interference}.

The differential cross section for process (\ref{eq:e+e-}) is
written in the following form
\begin{equation}
\mathrm{d}\sigma =\frac{1}{2s(2\pi )^{5}}\int \delta ^{4}(p_{1}+p_{2}-p_{-}-p_{+}-k)%
\frac{\mathrm{d}^{3}p_{+}}{2E_{+}}\frac{\mathrm{d}^{3}p_{-}}{2E_{-}}\frac{\mathrm{d}^{3}k}{2\omega}%
\overline{|M|^{2}},  \label{born}
\end{equation}
where $p_{+}=(E_{+},\mathbf{p}_{+})$, \ $p_{-}=(E_{-},\mathbf{p}_{-})$ and $%
k=(\omega=|\mathbf{k}|,\mathbf{k})$.

\subsection{Initial-state radiation}

\label{subsec:initial-state}

\hspace{0.5cm}Let us consider first the ISR contribution shown in
Fig.~1a. The amplitude squared can be written as
\begin{equation}
\overline{|M_{ISR}|^{2}}=-\frac{e^{6}}{q^{4}}|F_{\pi }(q^{2})|^{2}L_{\mu \nu
}^{(\gamma )}l^{\mu }l^{\nu },  \label{eq:ISR-squared}
\end{equation}
\begin{eqnarray*}
L_{\mu \nu }^{(\gamma )} &=&\Biggl[\frac{(q^{2}-t_{1})^{2}+(q^{2}-t_{2})^{2}%
}{t_{1}t_{2}}-\frac{2m_{e}^{2}q^{2}}{t_{1}^{2}}-\frac{2m_{e}^{2}q^{2}}{%
t_{2}^{2}}\Biggr]\tilde{g}_{\mu \nu } \\
&+&\Biggl(\frac{4q^{2}}{t_{1}t_{2}}-\frac{8m_{e}^{2}}{t_{1}^{2}}\Biggr)%
\tilde{p}_{2\mu }\tilde{p}_{2\nu }+\Biggl(\frac{4q^{2}}{t_{1}t_{2}}-\frac{%
8m_{e}^{2}}{t_{2}^{2}}\Biggr)\tilde{p}_{1\mu }\tilde{p}_{1\nu },\;
\end{eqnarray*}
where $\;\tilde{p}_{1\mu }=p_{1\mu }-q_{\mu }(p_{1}\cdot q)/q^{2}$ and
similarly for $\tilde{p}_{2\mu }$, and \ $\tilde{g}_{\mu \nu }=g_{\mu \nu
}-q_{\mu }q_{\nu }/q^{2}$.

If one integrates over the full, unrestricted phase space of the
final pions, the hadronic tensor can be integrated in the
invariant form (see \cite {Baier_65})
\begin{equation}
\frac{|F_{\pi }(q^{2})|^{2}}{16\pi ^{2}}\int \delta^4 (p_1 +p_2
-p_{+}-p_{-}-k)\ l_{\mu }l_{\nu } \frac{ \mathrm{d}^{3}p_{+}\mathrm{d}^{3}p_{-}}{E_{+}E_{-}} =%
\frac{q^{4}}{8\pi ^{2}\alpha ^{2}}\sigma (q^{2})\tilde{g}_{\mu \nu
}, \label{eq:Baier}
\end{equation}
which leads to the following cross section (``F'' stands for ``Full'')
\begin{equation}
\frac{\mathrm{d}\sigma^{(F)}_{ISR}}{\mathrm{d} \omega
\mathrm{d}\Omega}=\frac{\alpha \omega }{2\pi
^{2}s} \sigma (q^{2}) \biggl[ \frac{(q^{2}-t_{1})^{2}+(q^{2}-t_{2})^{2}}{%
t_{1}t_{2}}-\frac{2q^{2}m_{e}^{2}}{t_{1}^{2}}-\frac{2q^{2}m_{e}^{2}}{%
t_{2}^{2}} \biggr],
\end{equation}
where $\sigma (q^{2})=\displaystyle\frac{\mathstrut
\pi\zeta\alpha^2}{3q^2}|F_{\pi }(q^{2})|^{2}$ is the total cross
section for $e^{+} e^{-}\rightarrow \pi ^{+} \pi ^{-}$ (), $\zeta
=(1-{4m_{\pi }^{2}}/{q^{2}})^{3/2}$ and $\alpha =e^{2}/4\pi
\approx 1/137$ is
the fine-structure constant, $\mathrm{d}\Omega= \mathrm{d} \cos \theta \mathrm{d}\phi$ and $\theta$ ($%
\phi$) is the polar (azimuthal) angle for the emitted photon. Note
that the cross section does not depend on the azimuthal angle
$\phi$.

Integration over the photon angles leads to the result
\begin{equation}
\frac{\mathrm{d}\sigma_{ISR}^{(F)}}{\mathrm{d}q^2} = \sigma(q^2)
\frac{2 (s^2+q^4)}{\pi s^2(s-q^2)}(L-1), \; \; \;\;\;\;\;\;\;
L=\ln\frac{s}{m_e^2} . \label{eq:ISRF}
\end{equation}
In the last equation we used $\omega = (s-q^2)/{2 \sqrt{%
s}}$ and $\mathrm{d}q^2 = - 2\sqrt{s} \mathrm{d}\omega$. Note that
Eq.~(\ref{eq:ISRF}) holds for the full angular phase  space of photon;
another case of the restricted angular phase space of photon will be
studied elsewhere.

In some cases, due to experimental conditions, the entire phase space of
the pion is not available. Then it is not possible to use Eqs.~(\ref{eq:Baier})-(%
\ref{eq:ISRF}). In this situation one has to contract first the hadron and
lepton tensors and then carry out phase-space integration. From Eq.~(%
\ref{eq:ISR-squared}) we obtain the ISR contribution
\begin{equation}
\overline{|M_{ISR}|^{2}}=-\frac{4e^{6}}{q^{2}}|F_{\pi }(q^{2})|^{2}R,
\end{equation}
\begin{eqnarray*}
R &=&\frac{m_{\pi }^{2}}{q^{2}}F+\frac{\chi _{1}^{2}+\chi _{2}^{2}-\chi
_{1}(q^{2}-t_{2})-\chi _{2}(q^{2}-t_{1})}{t_{1}t_{2}} \\
&&-\frac{2m_{e}^{2}}{t_{2}^{2}}\Biggl(\frac{\chi _{1}}{q^{2}}-1\Biggr)-\frac{%
2m_{e}^{2}}{t_{1}^{2}}\Biggl(\frac{\chi _{2}}{q^{2}}-1\Biggr),
\end{eqnarray*}
where $\chi _{1,2}\equiv 2p_{1,2}\cdot p_{-}=\frac{1}{2}t_{1,2}-\frac{1}{2}
s-u_{1,2}$. Then the ISR cross section takes the form \cite{Khoze_02} (``R''
stands for ``Restricted'')
\begin{equation}
\mathrm{d}\sigma _{ISR}^{(R)}=\frac{12\sigma (q^{2})R}{s\zeta }\frac{\alpha }{4\pi ^{2}%
} \frac{\mathrm{d}^{3}k}{\omega }\frac{\mathrm{d}\varphi
_{+}}{2\pi }\frac{|\mathbf{p}
_{+}|\mathrm{d}E_{+}\mathrm{d}c_{+}}{E_{-}}\delta (2E-\omega
-E_{+}-E_{-}),\; \label{eq:ISR_full}
\end{equation}
where we introduced the electron (positron) energy $E=\sqrt{s}/2$
in the CM frame. Using the relation
\begin{equation}\label{eq:energy-conservation}
\int \frac{|\mathbf{p}
_{+}|\mathrm{d}E_{+}\mathrm{d}c_{+}}{E_{-}}\delta (2E-\omega
-E_{+}-E_{-}) \to \int \frac{|\mathbf{p} _{+}|^2 \mathrm{d}c_{+ }}{(2E-\omega)|%
\mathbf{p} _{+}|+\omega E_{+} \cos\theta_{\gamma +}}
\end{equation}
we obtain the corresponding ISR cross section
\begin{equation}
\frac{\mathrm{d}\sigma _{ISR}^{(R)}}{\mathrm{d}q^{2}}=\frac{3\sigma (q^{2})}{4E^{2}\zeta }%
\frac{\alpha }{2\pi }\frac{\omega }{E}\int\limits_{-1}^{1}\mathrm{d}\cos {\theta }%
\int\limits_{0}^{2\pi }\frac{\mathrm{d}\varphi _{+}}{2\pi }\int%
\limits_{-c_{m}}^{c_{max}}\frac{R|\mathbf{p}_{+}|^{2}\mathrm{d}c_+}{A |\mathbf{p}%
_{+}|+C E_{+}}.  \label{eq:RISR}
\end{equation}
If $\omega < 2E(E-m_\pi )/(2E-m_\pi)$ (which, for example,
corresponds to values $q^2 > 0.16$ GeV$^2$ at $\sqrt{s}=1$ GeV)
then we obtain
\begin{equation}\label{eq:E+}
E_+=\frac{2E(E-\omega)(2E-\omega)-\omega \cos\theta_{\gamma +}\sqrt{%
4E^2(E-\omega)^2- m_\pi^2[(2E-\omega )^2-\omega^2
\cos^2\theta_{\gamma +}]}}{(2E-\omega)^2-\omega^2
\cos^2\theta_{\gamma +}} .
\end{equation}
Here $c_+ =\cos\theta_+$, \ $\theta_+$ ($\varphi_+$) is the polar
(azimuthal) angle of the positively charged pion (we take $OZ$
axis along the vector $\mathbf{p_1}$), \ and $\cos \theta_{\gamma
+} = \mathbf{k} \mathbf{p_+} / |\mathbf{k}| |\mathbf{p_+}|$. In
this  energy region the angle $\theta_{\gamma +}$ can take
arbitrary values (for details see Appendix~E). Other notation can
be found in Ref.~\cite{Khoze_02}.

\subsection{Final-state radiation}

\label{subsec:final-state}

\hspace{0.5cm}The process of  photon radiation from the final
state is shown in Fig.~1b, where the dark rhomb denotes the set of
the contributing diagrams. The covariant decomposition of the FSR\
tensor $M_{f}^{\mu \nu }$ can be obtained from  Lorentz and
discrete symmetries (Appendix~A). This tensor involves three
gauge-invariant structures $\tau _{i}^{\mu \nu }$ and  invariant
functions $f_{i}$ $(i=1,2,3)$. The explicit form of the functions
$f_{i}$ in the framework of ChPT is discussed in Sect.~\ref
{sec:final-state-ChPT}. In terms of $f_{i}$ we obtain
\begin{eqnarray}
\overline{|M_{FSR}|^{2}} &=&\frac{e^{6}}{s^{2}}|F_{\pi }(s)|^{2}\biggl[%
a_{11}|f_{1}|^{2}+2a_{12}\mathrm{Re}(f_{1}f_{2}^{\ast
})+a_{22}|f_{2}|^{2} \nonumber
\\
&+&2a_{23}{Re}(f_{2}f_{3}^{\ast
})+a_{33}|f_{3}|^{2}+2a_{13}{Re}(f_{1}f_{3}^{\ast })\biggr],
\label{fsr}
\end{eqnarray}
\begin{eqnarray*}
a_{11} &=&\frac{1}{4}s(t_{1}^{2}+t_{2}^{2}),\;\;\;\;\;\;a_{33}=-\frac{s^{2}%
}{2}[t_{1}t_{2}l^{2}+2(u_{1}+u_{2})(u_{2}t_{1}+u_{1}t_{2})], \\
a_{22} &=&\frac{1}{8}\biggl(sl^{4}(t_{1}+t_{2})^{2}+4l^{2}\biggl[%
u_{1}{}^{2}(s^{2}+s(t_{1}+t_{2})+t_{2}^{2})+u_{2}{}^{2}(s^{2}+s(t_{1}+t_{2})+t_{1}^{2})
\\
&&+2u_{1}u_{2}[s^{2}+s(t_{1}+t_{2})-t_{1}t_{2}]\biggr]\biggr)%
+s(u_{1}^{2}+u_{2}^{2})(u_{1}+u_{2})^{2},
\end{eqnarray*}
\begin{eqnarray*}
a_{12} &=&\frac{1}{8}\biggl(%
sl^{2}(t_{1}+t_{2})^{2}+4u_{1}{}^{2}(s^{2}+st_{2}+t_{2}^{2})+4u_{2}{}^{2}(s^{2}+st_{1}+t_{1}^{2})
\\
&&+4u_{1}u_{2}[2s^{2}+s(t_{1}+t_{2})-2t_{1}t_{2}]\biggr), \\
a_{13} &=&\frac{s}{4}[%
(u_{1}+u_{2})(st_{1}+st_{2}+t_{1}t_{2})-u_{1}t_{2}^{2}-u_{2}t_{1}^{2}], \\
a_{23} &=&\frac{s}{4}\biggl[%
-l^{2}(u_{1}t_{2}^{2}+u_{2}t_{1}^{2}-(u_{1}+u_{2})t_{1}t_{2})-2s(u_{1}+u_{2})^{3}
\\
&&+2(u_{1}+u_{2})[u_{1}{}^{2}t_{2}+u_{2}{}^{2}t_{1}-u_{1}u_{2}(t_{1}+t_{2})]%
\biggr].
\end{eqnarray*}

In order to obtain the cross section $\mathrm{d}\sigma _{FSR}$ we
have to substitute Eq.~(\ref{fsr}) in Eq.~(\ref{born}) and
integrate over the phase space of the final particles.

In the case of the full phase space of pions, the integration can
be simplified using the method suggested in Ref.~\cite{Baier_65}.
In this case the squared matrix element  $\overline{|M_{FSR}|^2}$
in Eq.~(\ref{eq:total_amplitude}) can be integrated in the
invariant form
\begin{equation}
\int
\overline{|M_{FSR}|^2}\frac{\mathrm{d}^3p_-\mathrm{d}^3p_+}{E_-E_+}\delta^4(q-p_--p_+)=\frac{e^6}{s^2}
\overline{J_\mu J_\nu^*} W^{\mu\nu} .
\end{equation}

Taking into account the conditions $W^{\mu\nu}
Q_\mu=W^{\mu\nu}Q_\nu=0$, one can write $W^{\mu\nu}$ as
\begin{equation}\label{wmunu}
W^{\mu\nu}=h_1g^{\mu\nu}+\frac{Q^2}{(k \cdot
Q)^2}(h_1+Q^2h_2)k^\mu k^\nu+h_2Q^\mu
Q^\nu-\frac{h_1+Q^2h_2}{k\cdot Q}(k^\mu Q^\nu+k^\nu Q^\mu) ,
\end{equation}
where $h_{1,2}$ are functions of  $q^2$ and $Q^2$.

In the  framework of sQED $h_{1,2}$ were found in Ref.~\cite{Baier_65}
(see also Appendix~D). Using Eq.~(\ref{wmunu}) we obtain the
following equations for $h_{1,2}$ in terms of the functions $f_i$
for any model of FSR :
\begin{eqnarray}
h_2 (k\cdot q)^2
&=&\int\frac{\mathrm{d}^3p_+\mathrm{d}^3p_-}{E_+E_-}\delta^4(Q-p_+-p_--k)
[(k\cdot l)^2|f_2|^2+ (k\cdot Q)^2|f_3|^2  \nonumber \\
&&-2(k\cdot l) (k\cdot Q) Re(f_2 f_3^\star)][(k\cdot Q)^2
l^2+(k\cdot l)^2(Q^2-2k\cdot Q)],
\end{eqnarray}
\begin{eqnarray}
2h_1-h_2 Q^2 &=&
\int\frac{\mathrm{d}^3p_+\mathrm{d}^3p_-}{E_+E_-}\delta^4(Q-p_+-p_--k)
\{
2(k\cdot Q)^2|f_1|^2  \nonumber \\
&& + [ l^4 (k\cdot Q)^2+l^2(k\cdot l)^2 (Q^2-2k\cdot Q)+2(k\cdot
l)^4]
|f_2|^2  \nonumber \\
&& +2[(k\cdot l)^2 Q^2+l^2 (k\cdot Q)^2]Re(f_1 f_2^\star)-4Q^2
(k\cdot Q)
(k\cdot l)Re(f_1 f_3^\star)  \nonumber \\
&& -4 Q^2 (k\cdot l)^3 Re(f_2 f_3^\star)+[(k\cdot l)^2
Q^4+2(k\cdot
l)^2 Q^2 (k\cdot Q)  \nonumber \\
&& -l^2 Q^2 (k\cdot Q)^2]|f_3|^2 \}
\end{eqnarray}
with $Q^2=s$. In Appendix D the explicit form of $%
h_{1,2}$ is presented in the framework of ChPT.

Then the cross section $\mathrm{d}\sigma_{FSR}^{(F)}/\mathrm{d}
\omega \mathrm{d} cos\theta $ takes the form
\begin{equation}  \label{sectFSR}
\frac{\mathrm{d}\sigma_{FSR}^{(F)}}{\mathrm{d} \omega \mathrm{d} cos\theta }=\frac{\alpha^3\omega}{%
4\pi s^2} [h_1-\frac{t_1t_2}{2s\omega^2}(h_1+s h_2) ].
\end{equation}
Integrating this equation over the polar angle of the emitted
photon we find
\begin{equation}\label{FSR_fin}
\frac{\mathrm{d}\sigma_{FSR}^{(F)}}{\mathrm{d}q^2}=\frac{\alpha^3(s-q^2)}{24
\pi s^3}(2h_1-s h_2).
\end{equation}

If we deal with the restricted phase space we can use the same
arguments which led to Eq.~(\ref{eq:RISR}) in
Sect.~\ref{subsec:initial-state}. Then the cross section can be
written as
\begin{equation}
\frac{\mathrm{d}\sigma_{FSR}^{(R)}}{\mathrm{d}q^2} =
\frac{\overline{|M_{FSR}|^2}} {32s(2\pi)^3
}\frac{\omega}{\sqrt{s}}\frac{|\mathbf{p_+}|^2}{(2E-\omega)
|\mathbf{p_+}|+\omega E_+ c_{\gamma +}} \frac{\mathrm{d}\varphi_+}{2\pi}%
\mathrm{d}c_+\mathrm{d}\cos\theta ,  \label{eq:RFSR}
\end{equation}
where 
$\overline{|M_{FSR}|^2}$ is determined in Eq.~(\ref{fsr}).

\subsection{Interference}

\label{subsec:interference}

\hspace{0.5cm}The interference part of the squared invariant
amplitude $\overline{|M|^{2}}$ is written in terms of the
invariant functions $f_{i}$ and the pion form factor as follows
\begin{equation}
\mathrm{Re}(\overline{M_{ISR}M_{FSR}^{\ast }})=-\frac{e^{6}}{4sq^{2}}\biggl[%
A_{1}\mathrm{Re}(F_{\pi }(q^{2})f_{1}^{\ast
})+A_{2}\mathrm{Re}(F_{\pi }(q^{2})f_{2}^{\ast
})+A_{3}\mathrm{Re}(F_{\pi }(q^{2})f_{3}^{\ast })\biggr],
\end{equation}
where the coefficients $A_{i}$ are
\begin{eqnarray}\label{A_i}
A_{1} &=&-2u_{1}(\frac{-s^{2}+(t_{1}-t_{2})s+t_{1}t_{2}}{t_{1}}+\frac{%
s^{2}+t_{2}^{2}}{t_{2}})+2u_{2}(\frac{-s^{2}+(t_{2}-t_{1})s+t_{1}t_{2}}{t_{2}%
}+\frac{s^{2}+t_{1}^{2}}{t_{1}}), \nonumber \\
A_{2} &=&-4\frac{u_{1}{}^{2}+u_{2}{}^{2}}{t_{1}t_{2}}\biggl[\bigl(%
s(t_{1}-t_{2})-t_{2}(t_{1}+t_{2})\bigr)u_{1}-\bigl(%
s(t_{2}-t_{1})-t_{1}(t_{1}+t_{2})\bigr)u_{2}\biggr] \nonumber \\
&-&\frac{l^{2}}{t_{1}t_{2}}\biggl(%
u_{1}[2s^{2}(t_{1}-t_{2})+s(t_{1}+t_{2})(t_{1}-3t_{2})-t_{2}(t_{1}-t_{2})^{2}]
\nonumber\\
&-&u_{2}[2s^{2}(t_{2}-t_{1})+s(t_{1}+t_{2})(t_{2}-3t_{1})-t_{1}(t_{2}-t_{1})^{2}] %
\biggr), \nonumber\\
A_{3} &=&-2s(t_{1}-t_{2})l^{2}-4s\biggl(u_{1}{}^{2}(2+\frac{s+t_{2}}{t_{1}}
)-u_{2}{}^{2}(2+\frac{s+t_{1}}{t_{2}})+\frac{s(t_{2}-t_{1})}{t_{1}t_{2}}
u_{1}u_{2}\biggr).
\end{eqnarray}

We would like to mention that the cross sections $\mathrm{d}\sigma _{ISR}$ and $%
\mathrm{d}\sigma _{FSR}$ are symmetric under the interchange of
$\pi ^{+}$ and $\pi ^{-}$ momenta, while the interference term
$\mathrm{d}\sigma _{IFR}$ is antisymmetric:
\begin{eqnarray}
\mathrm{d}\sigma _{ISR}(p_{+},p_{-}) &=&\mathrm{d}\sigma
_{ISR}(p_{-},p_{+}),\ \; \ \ \ \ \ \
\mathrm{d}\sigma _{FSR}(p_{+},p_{-})=\mathrm{d}\sigma _{FSR}(p_{-},p_{+})  \nonumber \\
\mathrm{d}\sigma _{IFR}(p_{+},p_{-}) &=&- \mathrm{d}\sigma
_{IFR}(p_{-},p_{+}). \label{eq:p+p-interchange}
\end{eqnarray}
Therefore $\mathrm{d}\sigma _{IFR}$ integrated over the symmetric
phase space of the pions (for example, for the full unrestricted
phase space) is equal to zero. Other implications of
Eqs.~(\ref{eq:p+p-interchange}) are considered in the next
sections.

For the restricted phase space of pions we have a result
analogous to (\ref{eq:RFSR})
\begin{equation}  \label{interf}
\frac{\mathrm{d}\sigma_{IFR}^{(R)}}{\mathrm{d}q^2} = \frac{\mathrm{Re}(\overline{M_{ISR}M_{FSR}^{\ast }}%
)} {16s(2\pi)^3 }\frac{\omega}{\sqrt{s}}\frac{|\mathbf{p_+}|^2}{%
(2E-\omega) |\mathbf{p_+}|+\omega E_+ c_{\gamma +}} \frac{\mathrm{d}\varphi_+}{%
2\pi}\mathrm{d}c_+\mathrm{d}\cos\theta .
\end{equation}

\section{Final-state radiation in the framework of ChPT}
\label{sec:final-state-ChPT}

\hspace{0.5cm}Based on results of Appendix~A we can write the FSR
tensor $M_{f}^{\mu \nu }$ in the form
\begin{equation}
M_{F}^{\mu \nu }=f_{1}\tau _{1}^{\mu \nu }+f_{2}\tau _{2}^{\mu \nu
}+f_{3}\tau _{3}^{\mu \nu }.  \label{eq:tensor-MF}
\end{equation}
In framework of ChPT with vector and axial-vector mesons
\cite{Ecker_89} (see details in Appendix~B) the process $\gamma
^{\ast }\rightarrow \pi ^{+}\pi ^{-}\gamma $ is described at
tree level by the diagrams shown in Fig.~2.
\begin{figure}[tbp]
\begin{center}
\epsfig{file=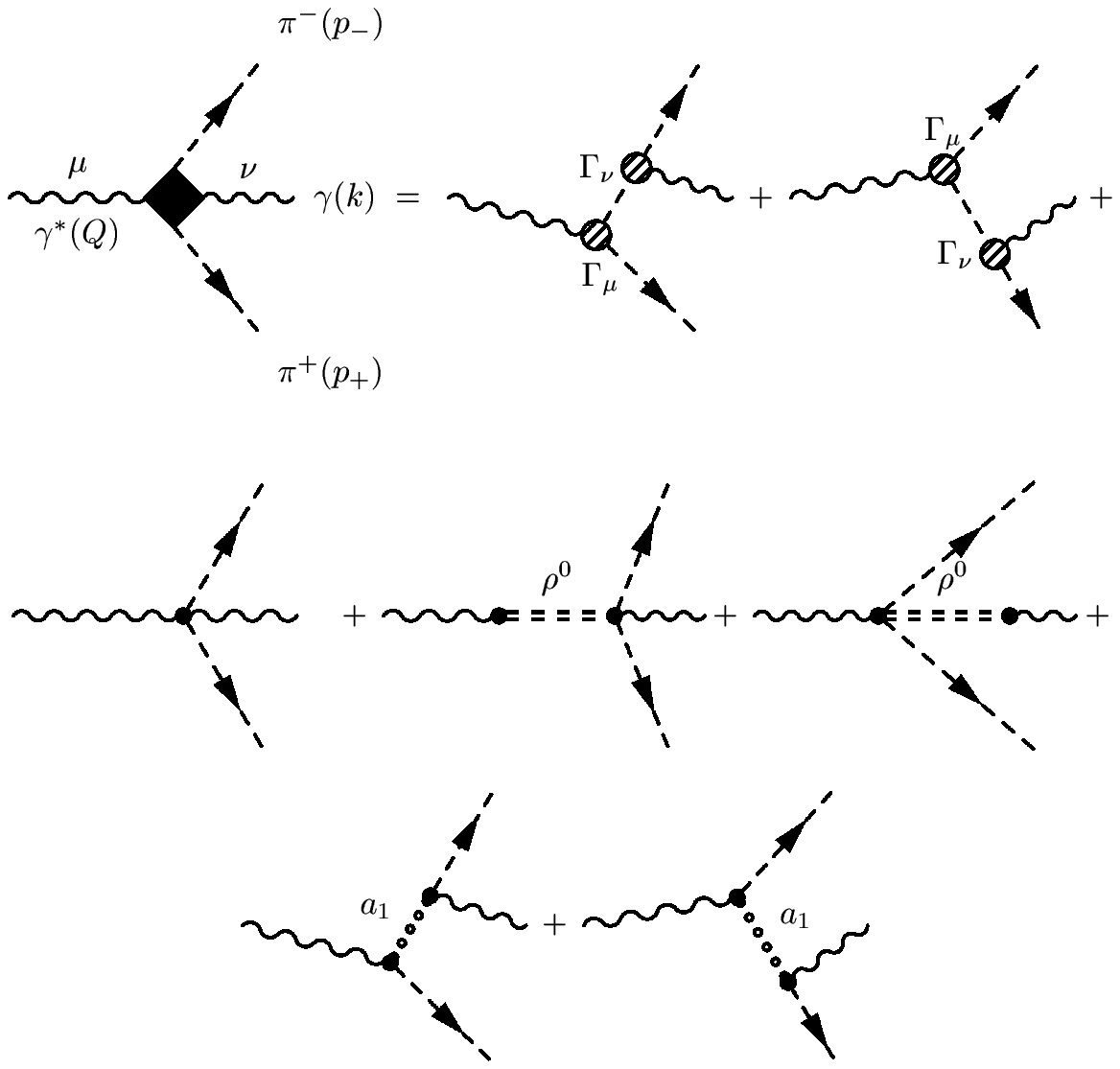,width=9cm,height=10cm}
\vspace{0.5cm}

\epsfig{file=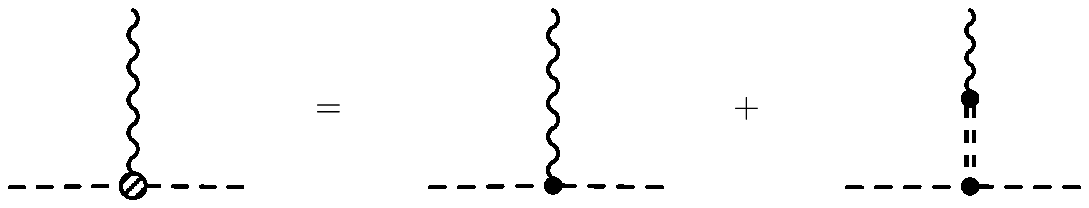,width=7cm,height=2cm} \label{fig2}
\end{center}
\caption{Diagrams for FSR in the framework of ChPT. Dashed lines
depict pion, wavy lines -- photon, double-dashed lines --
$\rho^0-$meson, and dotted lines -- $a_1-$meson. The hatched
circles denote the irreducible $\gamma \pi \pi $ vertex }
\end{figure}
Using results from Appendix B we find the invariant functions $f_{i}\equiv
f_{i}(Q^{2},k\cdot Q,k\cdot l)$
\begin{equation}
f_{i}=f_{i}^{sQED}+\Delta f_{i},  \label{eq:functions-fi}
\end{equation}
where $f_{i}^{sQED}$ correspond to  sQED
\begin{eqnarray}
f_{1}^{sQED}&=&\frac{2k\cdot Q F_{\pi }(Q^2)}{(k\cdot
Q)^{2}-(k\cdot l)^{2}},
\nonumber \\
f_{2}^{sQED}&=&-\frac{2F_{\pi }(Q^2)}{(k\cdot Q)^{2}-(k\cdot l)^{2}},\,\ \ \
\ \ \ \ \ \ \ f_{3}^{sQED}=0,  \label{eq:functions-in-sQED}
\end{eqnarray}
and $\Delta f_{i}$ are additional contributions
\begin{eqnarray}
\Delta f_{1} &=&\frac{F_{V}^{2}-2F_{V}G_{V}}{f_{\pi }^{2}}\biggl(\frac{1}{%
m_{\rho }^{2}}+\frac{1}{m_{\rho }^{2}- Q^2}\biggr)  \nonumber \\
&-&\frac{F_{A}^{2}}{f_{\pi }^{2}m_{a}^{2}}\biggl[ 2+\frac{(k\cdot l)^{2}}{%
D(l)D(-l)}+\frac{(Q^{2}+k\cdot Q)[4m_{a}^{2}-(Q^{2}+l^{2}+2k\cdot Q)] }{
8D(l)D(-l)}\biggr],  \label{eq:delta-f1} \\
\Delta f_{2} &=&-\frac{F_{A}^{2}}{f_{\pi }^{2}m_{a}^{2}}\frac{%
4m_{a}^{2}-(Q^{2}+l^{2}+2k\cdot Q)}{8D(l)D(-l)},  \label{eq:delta-f2} \\
\Delta f_{3} &=&\frac{F_{A}^{2}}{f_{\pi }^{2}m_{a}^{2}}\frac{k\cdot l}{%
2D(l)D(-l)}.  \label{eq:functions-in-ChPT}
\end{eqnarray}
Here $D(l)=m_{a}^{2}-(Q^{2}+l^{2}+2k\cdot Q+4k\cdot l)/4$. The
functions $\Delta f_{i}$ and $f_{i}^{sQED}$ obey the same symmetry
relations as given by Eq.~(\ref{eq:A5}) for functions $f_i$. The
EM form factor in Eqs.~(\ref{eq:functions-in-sQED}) for the
on-shell pion follows from Eq.~(\ref{eq:B5}):
\begin{equation}
F_{\pi }(Q^{2})=1+\frac{F_{V}G_{V}}{f_{\pi }^{2}}\frac{Q^2}{m_{\rho
}^{2}-Q^{2}}. \label{eq:pion-form-factor}
\end{equation}
To account for the finite width of the vector meson one can
substitute in Eqs.~(\ref{eq:delta-f1}) and
(\ref{eq:pion-form-factor})
\begin{equation}
m_{\rho }^{2}-Q^{2} \rightarrow  m_{\rho }^{2}-Q^{2}-im_{\rho
}\Gamma _{\rho }(Q^{2}), \quad \quad \quad \Gamma_{\rho }(Q^{2}) =
\frac{m_\rho G_V^2 }{48 \pi f_\pi^4}{Q^{2}}
(1-\frac{4m_\pi^2}{Q^{2}} )^{3/2} \theta (Q^{2}-4m_{\pi }^{2}),
\end{equation}
where $\Gamma _{\rho }(Q^{2})$ is the energy-dependent width for
the $\rho \rightarrow \pi \pi $ decay.  In a similar way one can
include in Eqs.~(\ref{eq:delta-f1})--(\ref{eq:functions-in-ChPT})
the width of the decay $a_{1} \rightarrow 3 \pi$. The analytical
form of $\Gamma _{a_1}(Q^{2})$ can be taken from, \textit{e.g.},
Ref.~\cite{Ecker_02}.

Using the form of $f_{1,2,3}$ we can find the functions $h_{1,2}$
from Sect.~\ref{subsec:final-state} appearing in the FSR cross
section. The expressions are rather lengthy and are listed in
Appendix~D.

We would like to mention that the Compton $\gamma\pi\to\gamma\pi$ scattering amp litude in the framework of ChPT \cite{Ecker_89} was calculated in Ref.~\cite{Donoghue_93}. Having compared Eqs.~(37)-(40) of
Ref.~\cite{Donoghue_93} with 
Eqs.~(\ref{eq:delta-f1})-(\ref{eq:functions-in-ChPT}) of this paper, we
have concluded that the $\rho$-meson contributions are equal whereas the
$a_1$-meson contributions are different.

 
In addition to the even-intrinsic-parity contributions considered
above there is an odd-intrinsic-parity part. The corresponding
Lagrangian \cite{Pich_90,Ruiz_03} describes processes which do not
conserve intrinsic parity, such as $\rho \to \pi \gamma$. The
contribution of the two-step mechanism $\gamma^* \to \rho^\pm
\pi^\mp \to \pi^+ \pi^- \gamma$ to the FSR tensor is evaluated in
Appendix~F.

\section{Results of calculation}
\label{sec:results}

\hspace{0.5cm} Table 1 lists the parameters of the model. The
couplings $F_V, F_A$ are determined from the experimental decay
widths \cite{PDG}: $\Gamma (\rho^0 \to e^+ e^- )=6.85 \pm 0.11$
keV and $\Gamma (a_1 \to \pi \gamma )=640 \pm 246$ keV, while
$G_V$ is fixed from the width $\Gamma ({\rho \to \pi \pi})=150.7
\pm 2.9$ MeV (we neglect the chiral corrections here). The pion
weak-decay constant is $f_{\pi }=92.4$ MeV.

\begin{table}
\begin{center}
\begin{tabular}{|c|c|c|c|c|}
\hline
meson & m (GeV) & $G_V$ (GeV) & $F_V$ (GeV) & $F_A$ (GeV) \\
\hline
$\rho$ & 0.775 & 0.066 & 0.14 & -- \\
 $a_1$ & 1.23 & -- & -- & 0.122 \\
\hline
\end{tabular}
\end{center}
\caption{Masses and coupling parameters of vector and axial-vector
mesons } \label{tab:mesons}
\end{table}

\subsection{Charge asymmetry}
\label{subsec:charge_asymmetry}

\hspace{0.5cm}We will illustrate the  results obtained in the
previous sections by considering the charge asymmetry
\cite{Czyz_03}
\begin{equation}
A=\frac{N(\theta _{+})-N(\theta _{-})}{N(\theta _{+})+N(\theta
_{-})} \label{eq:asymmetry-1}
\end{equation}
for ``collinear'' kinematics in which the hard photon is radiated
inside a narrow cone with the opening angle $2\theta _0$,
$\theta_0\ll 1$, along the direction of initial electron. We choose $\theta_0=7^o$. In these
conditions the asymmetry takes the form
\begin{equation}
A \approx \frac{\mathrm{d}\sigma
_{IFR}^{(R)}}{\mathrm{d}q^2\mathrm{d}c_{+}} \bigg [
\frac{\mathrm{d}\sigma _{ISR}^{(R)}}{\mathrm{d}q^2\mathrm{d}c_{+}}
\bigg]^{-1}, \label{eq:asymmetry-2}
\end{equation}
where we neglected the FSR contribution compared to the ISR one in
the denominator.

The ISR cross section for collinear kinematics was obtained in
Ref.~\cite{Khoze_02}. We use Eqs.~(26)--(30) of
Ref.~\cite{Khoze_02}, which were derived in the
quasi--real--electron approximation. It is convenient to rewrite
these results as follows
\begin{equation}
\frac{\mathrm{d}\sigma
_{ISR}^{(R)}}{\mathrm{d}q^2\mathrm{d}c_{+}}=\frac{\pi \alpha ^2\xi
|F_\pi (q^2)|^2}{3q^2s}\frac \alpha {2\pi }P(z,L_0)A(q^2,c_{+})
\end{equation}
$$
P(z,L_0)=\frac{s^2+q^4}{s(s-q^2)}L_0-\frac{2q^2}{s-q^2},\;\;\;
\; \; L_0=\ln \frac{%
s\theta _0^2}{4m_e^2},
$$
$$A=\frac{12}\xi \frac{z[(1+z)K-(1-z)c_{+}]^2U}{%
K[(1+z)^2-(1-z)^2c_{+}^2]^2}, \;\;\; \; \; U=\frac{\chi
_1}s-\frac{\chi_1^2}{s^2}-\frac{m_\pi ^2}{q^2},\;\;\; \;
z=\frac{q^2}{s} ,$$
$$\frac{\chi
_1}s=\frac{z[1+z-2Kc_{+}+(1-z)c_{+}^2]}{(1+z)^2-(1-z)^2c_{+}^2}%
,\;\;\; \; \; K=\sqrt{1-\frac{m_\pi
^2}{sz^2}[(1+z)^2-(1-z)^2c_{+}^2]}.
$$

In order to obtain $\mathrm{d}\sigma
_{IFR}^{(R)}/\mathrm{d}q^2\mathrm{d}c_{+}$
we should integrate the right--hand side of Eq.~(\ref{interf})
over $\varphi_+$ and $\theta$. Since the right--hand side has no
singularity at the point $\theta=0$ the integration
over $\varphi_+$ and $\theta $ can be easily done  numerically.
\begin{figure}[tbp]
\label{fig_as}
\begin{center}
\epsfig{file=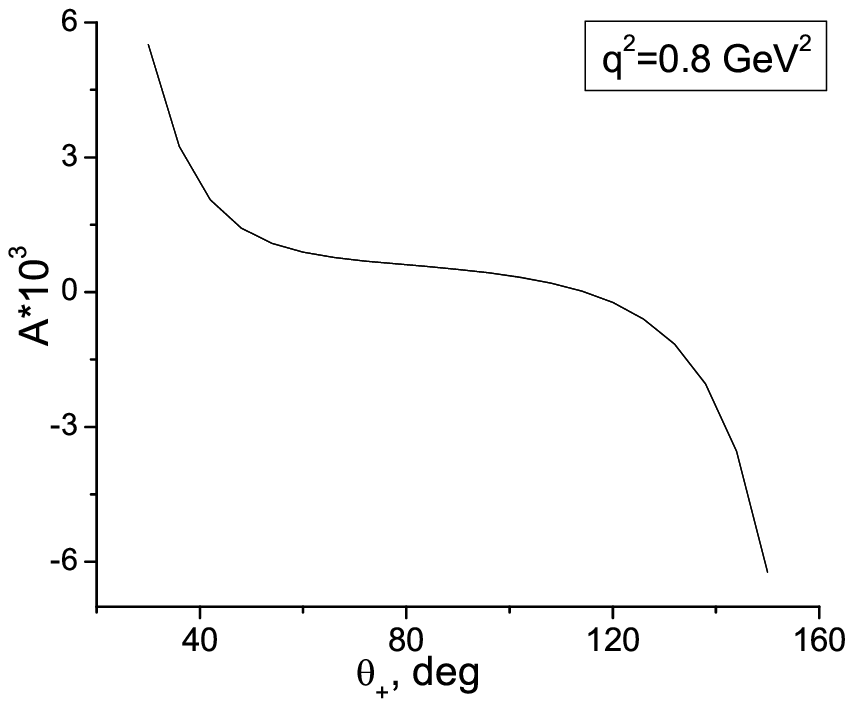,width=5.3cm,height=4.5cm} 
\epsfig{file=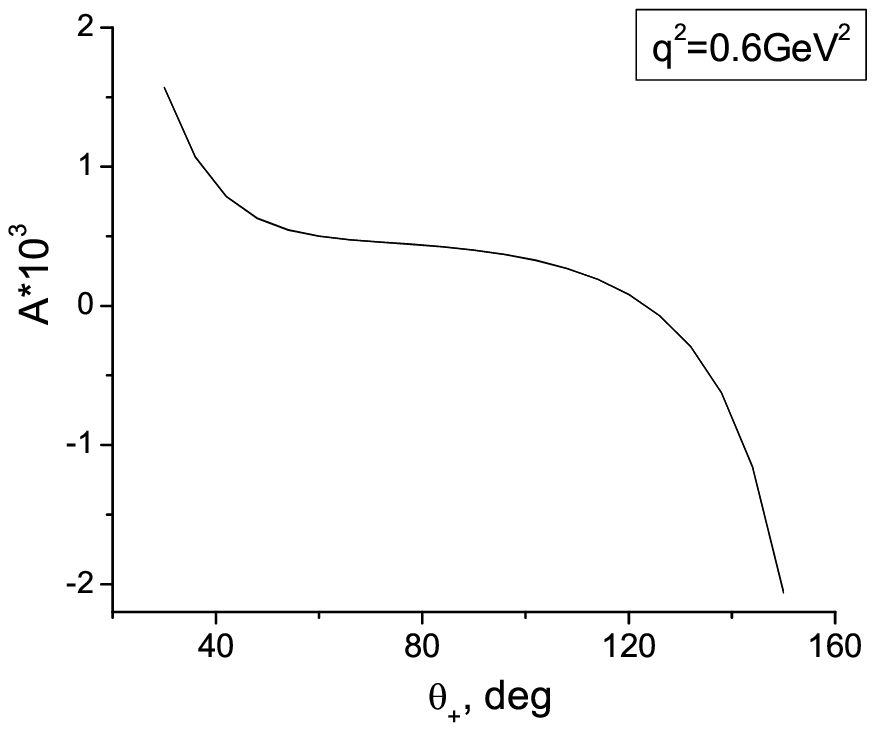,width=5.3cm,height=4.5cm}
\epsfig{file=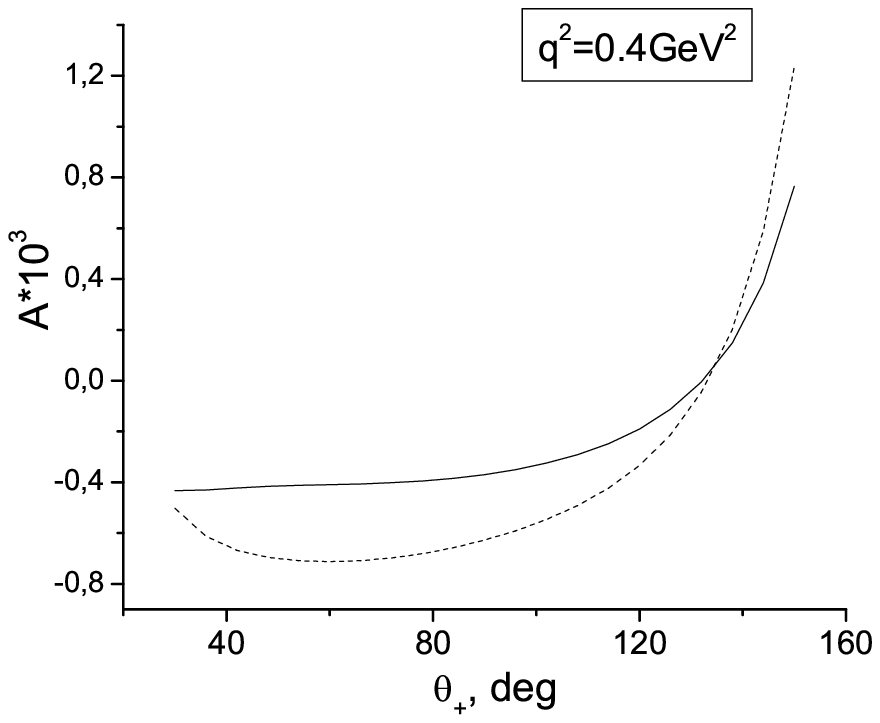,width=5.3cm,height=4.5cm}
\end{center}
\vspace{-1cm} \caption{Charge asymmetry as a function of pion
polar angle at fixed $q^2$ for $s=1$ GeV$^2$. Here the solid line
corresponds to sQED, the dashed line -- the full result in ChPT. }
\end{figure}
\begin{figure}[t]
\label{as_theta}
\begin{center}
\epsfig{file=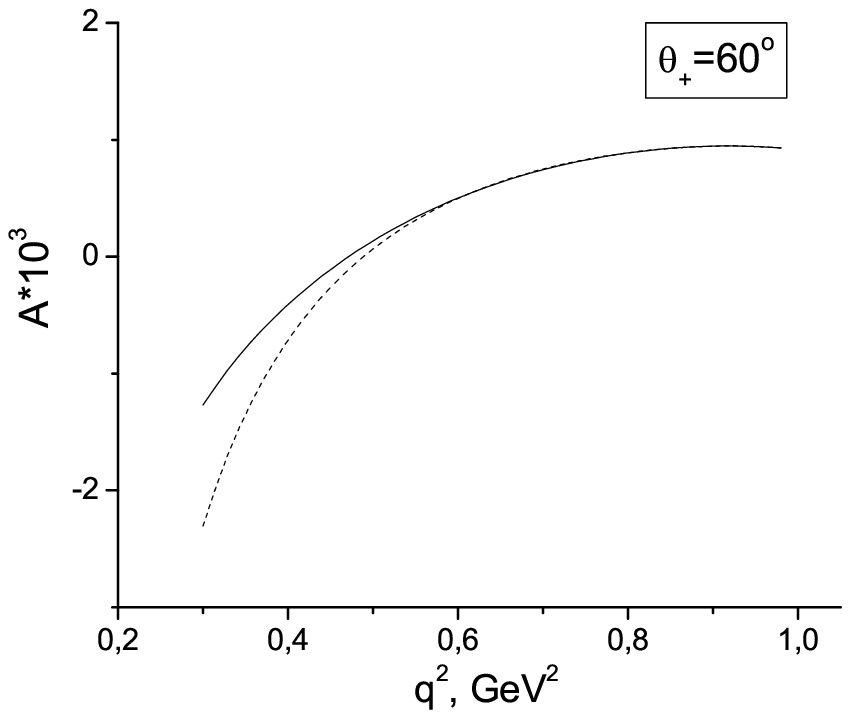,width=5.3cm,height=4.5cm} 
\epsfig{file=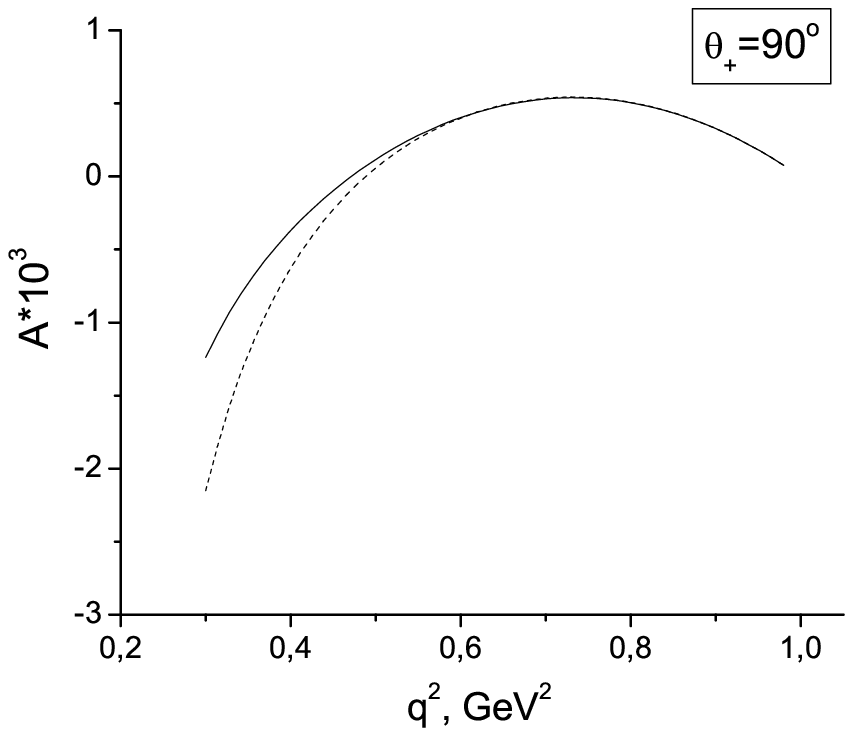,width=5.3cm,height=4.5cm}
\epsfig{file=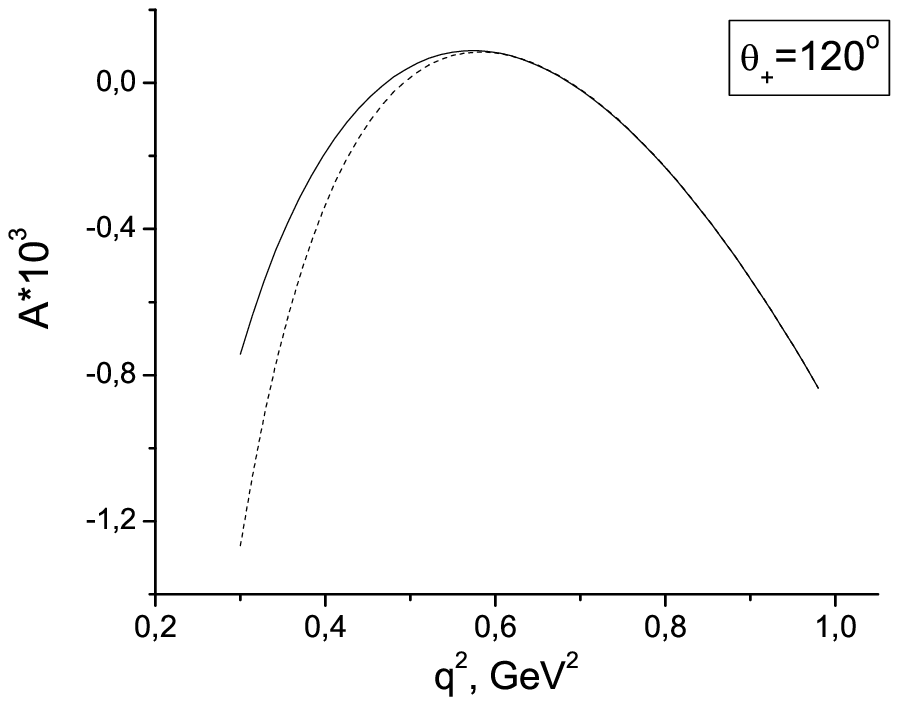,width=5.3cm,height=4.5cm}
\end{center}
\vspace{-1cm} \caption{Charge asymmetry as a function of $q^2$ at
fixed pion polar angle for $s=1$ GeV$^2$. Notation for the curves
is the same as in Fig.~3.}
\end{figure}

In Figs.~3 and 4 we show the asymmetry dependence on pion polar
angle (at fixed $q^2$) and on $q^2$ (at fixed pion polar angle).
It follows from the calculations that the asymmetry changes sign
at about $q^2=0.5$ GeV$^2$ (see Fig.~4). At all pion angles the
difference between sQED and ChPT shows up only at small values of
$q^2$, i.e at the high photon energies.

The sQED description is adequate for soft photon emission. It
follows from
Eqs.~(\ref{eq:functions-in-sQED})--(\ref{eq:functions-in-ChPT})
that at small photon energies, $f_{1}^{sQED}\sim\frac{\mathstrut
1}{\omega}$, \ $f_{2}^{sQED}\sim\frac{\mathstrut 1}{\omega^2}$,
whereas $\Delta f_i$ which are responsible for the deviation from
sQED behave rather as constants. Only at large photon energies the
contribution from the intermediate $a_1$--meson (the last two
diagrams in Fig.~2) can be sizable, because the denominators
$D(l)D(-l)$ in
Eqs.~(\ref{eq:delta-f1})--(\ref{eq:functions-in-ChPT}) approach
the $a_1$--meson pole with the photon energy increasing.

For high value of $q^2$  the difference between predictions of
sQED and full calculation in ChPT is small. For example, at
$q^2=0.8$ GeV$^2$ and $q^2=0.6$ GeV$^2$ it is less than $1\%$ (the
dashed and solid lines in Figs.~3 and 4 almost coincide). Taking
into account that the asymmetry itself is less than $10^{-2}$, the
experimental observation of such deviations in the energy region
$q^2\geq 0.6$ GeV$^2$ is problematic.

Additional contribution coming from the process $\gamma^* \to
\rho^\pm \pi^\mp \to \pi^+ \pi^- \gamma$ turns out very small (see
Appendix~F) and does not change the above conclusion.

In order to test the calculation we can check that the asymmetry,
integrated over the symmetric phase space of the pions,  vanishes.
Since no restriction has been imposed on the $\pi^-$ polar angle
we impose no restriction on the $\pi^+$ polar angle, i.e. choose \
$0^\circ \leq \theta_+,\theta_- \leq 180^\circ$. Indeed, the
integrated asymmetry is equal to zero.

\subsection{Contribution from $e^+ e^- \to \pi^+ \pi^- \gamma$ to AMM
of the muon } \label{subsec:AMM_muon}

\begin{figure}[t]
\label{muon_anom}
\begin{center}
\epsfig{file=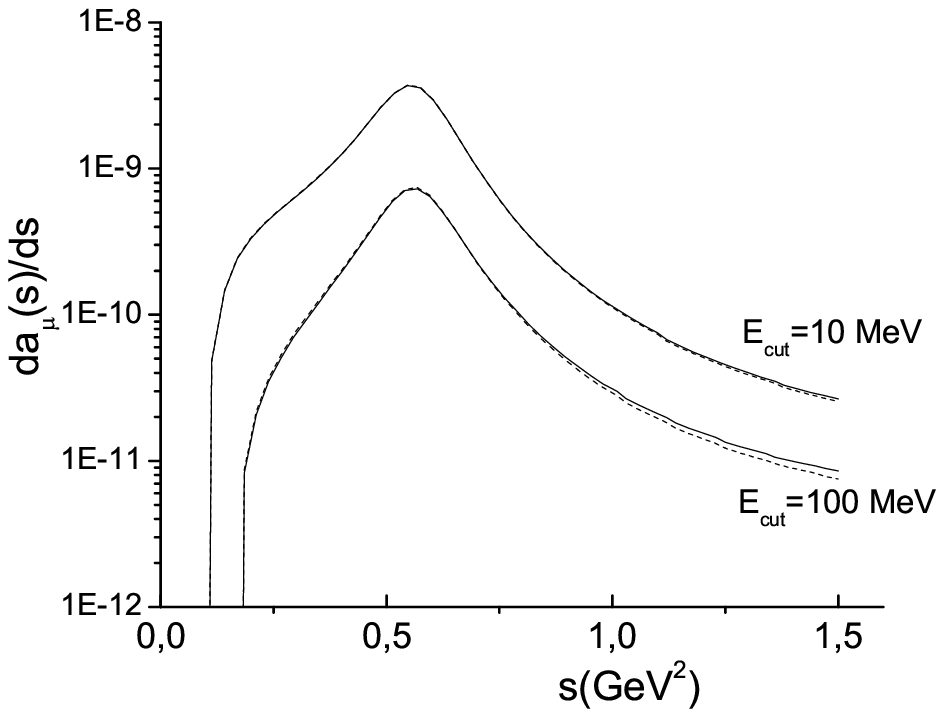,width=6.3cm,height=5.5cm} 
\epsfig{file=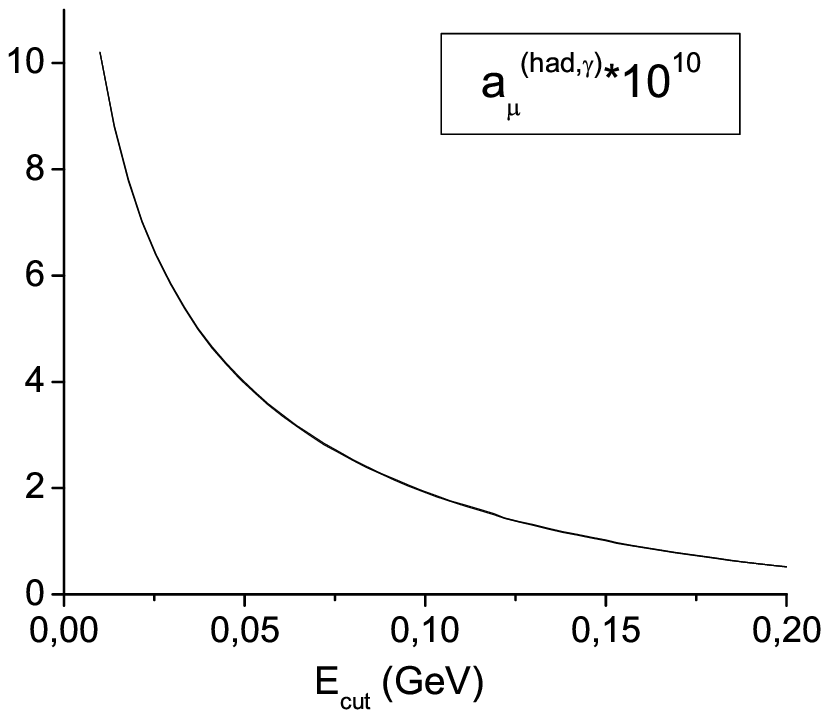,width=6.3cm,height=5.5cm}
\end{center}
\vspace{-1cm} \caption{Differential contribution to
$a_\mu^{(had,\gamma)}$ from $e^+e^-\to\pi^+\pi^-\gamma$, where
$\gamma$ is the hard photon with energy $\omega\geq E_{cut}$ (left
panel). Integrated contribution to $a_\mu^{(had,\gamma)}$ as a
function of $E_{cut}$ for $s_{max}=1.5$ GeV$^2$ (right panel). The
solid (dashed) line corresponds to the sQED (full) result.}
\end{figure}

\hspace{0.5cm} Now we apply  the previous results to the calculation of
$a_\mu^{had,\gamma}$ which arises from the $\pi^+ \pi^- \gamma$
channel. We should mention that only the radiation of the hard
photon with the energy $\omega\geq E_{cut}$ is taken into account.

According to  \cite{Brodsky_68} $a_\mu^{had,\gamma}$ can be
written in terms of the dispersion integral

\begin{eqnarray}\label{a_mu}
&& a_\mu^{had,\gamma}=\frac{\alpha^2}{3\pi^2
}\int_{4m_\pi^2}^{\infty}R^\gamma(s)K(s)\frac{\mathrm{d} s}{s} ,
\; \; \; \; \; \; \;
 R^\gamma(s)=\frac{\sigma^{\pi^+\pi^-,\gamma}(s)}{\sigma^{\mu^+\mu^-}(s)} , \label{s_int} \\
&&
\sigma^{\mu^+\mu^-}(s)=\frac{4\pi\alpha^2}{3s}\Bigl(1+\frac{2m_\mu^2}{s}\Bigr)
\sqrt{1-\frac{4m_\mu^2}{s}} , \nonumber\\
&& K(s)=\int_0^1\frac{x^2(1-x)}{x^2+(1-x)s/m_\mu^2}\mathrm{d}x,
\nonumber
\end{eqnarray}
where $m_\mu$ is the muon mass.

To obtain the cross section $\sigma^{\pi^+\pi^-,\gamma}(s)$ we
have to integrate Eq.~(\ref{FSR_fin}) over $q^2$ from $4m_\pi^2$
to $q^2_{max}$. The value of $q^2_{max}$ is determined from the
equality $q^2_{max}=s-2\sqrt{s}E_{cut}$. From the condition
$q^2_{max}\geq 4m_\pi^2$, the lower limit of  the integration  in
Eq.~(\ref{s_int}) is found to be
$s_{min}=(E_{cut}+\sqrt{4m_\pi^2+E_{cut}^2})^2$. The upper
limit in (\ref{s_int}) is replaced by a finite
$s_{max}$. The value of $s_{max}$ is chosen $s_{max} = 1.5$
GeV$^2$, which is about of $O(m_{a_1}^2)$, the upper limit of the
applicability of ChPT with $\rho$ and $a_1$ mesons. The dependence
of $a_\mu^{had,\gamma}$ on energy $E_{cut}$ is shown in Fig.~5.

\begin{figure}[t]
\begin{center}
\epsfig{file=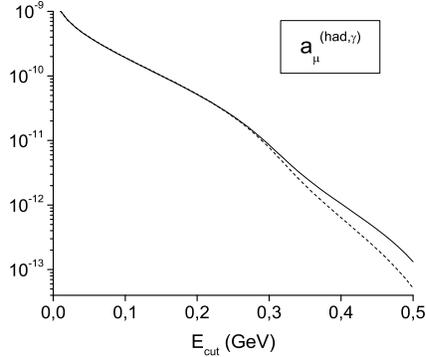,width=6.3cm,height=5.5cm}
\end{center}
\vspace{-1cm} \caption{Integrated contribution to
$a_\mu^{(had,\gamma)}$ as a function of $E_{cut}$ for
$s_{max}=1.5$ GeV$^2$. Notation for the curves is the same as in
Fig.~3.}\label{acut_500}
\end{figure}

As follows from our calculations the additional contributions to
$a_\mu^{had,\gamma}$ stemming from ChPT are very small compared to
the sQED result. This is in line with the conclusion of
Ref.~\cite{Dubinsky}. Even for relatively large cut-off energy
$E_{cut}=200$ MeV the full result in ChPT differs from the sQED
result only by $3.5\%$. For that reason the solid and dashed lines
in Fig.~5 almost coincide. At the same time with increasing 
photon energy sQED begins to loose its predictive power.  This is
demonstrated in Fig.~6. In this region of energies the
contribution from the $a_1$--meson is considerable and has to be
taken into account. For example, at a  photon energy about 500
MeV the deviation from sQED reaches $60\%$. However such a
deviation (which is of the order of $10^{-12}$) is beyond the
accuracy of the present measurements of the muon AMM.

\section{Conclusions}\label{sec:conclusions}

\hspace{0.5cm} In this article the  FSR of a hard
photon in the $e^+e^-\to\pi^+\pi^-\gamma$ reaction has been
considered in framework of ChPT with vector $\rho$ and
axial--vector $a_1$ mesons. The respective Lagrangian generates
effective $\mathcal{O}(p^{4})$ chiral terms and, as substantiated
in \cite{Ecker_89}, is adequate for description of processes at
energies up to about 1 GeV.

Our consideration of FSR is motivated by the necessity to study
the model dependence of the hadronic contribution
$a_\mu^{had,\gamma}$ to AMM of the muon. We have demonstrated
that this dependence is weak, in particular, in the region of
energies up to $s=1.5$ GeV$^2$ the differences between predictions
of ChPT and sQED are very small compared to the present
experimental accuracy. In general, the deviation of ChPT from sQED
increases with increasing the minimal photon energy $E_{cut}$.
However this deviation becomes essential only if the energy of the
photon exceeds 400 MeV, in the region where $a_\mu^{had,\gamma}$
itself is beyond the existing experimental precision. To observe
such effects the experimental accuracy has to be increased by at
least one order of magnitude.

In fact, this small deviation is not surprising. Firstly, at fixed
value of $s$ the low--energy photon region is described similarly
in both models and, as follows from the calculation, this region
dominates in $a_\mu^{had,\gamma}$. Secondly, the main contribution
to the integral over $s$ in Eq.~(\ref{a_mu}) comes from the region
of the $\rho$--resonance, which is accounted for in the same way
in sQED and ChPT through the VMD model. Therefore, the integral
$a_\mu^{had,\gamma}$ is not sensitive to the chiral dynamics (see
also discussion in \cite{Dubinsky}).

The developed approach has also allowed us to investigate the
$C$--odd asymmetry of the cross section caused by the ISR -- FSR
interference. In general, measurements of the asymmetry can test
the FSR amplitude. We considered radiation of the photon at the
small angle relative to the direction of the electron momentum,
$\theta<\theta_0=15^\circ$. In these conditions the absolute value
of asymmetry is  of the order of $\theta_0^2\sim
10^{-2}$. According to our calculations the difference between
sQED and ChPT shows up only at the high photon energies
$\omega\geq 0.3$ GeV. For the smaller photon energies the
difference is less than $1\%$. Since the asymmetry itself is less
than $10^{-2}$ for the selected collinear kinematics the
experimental observation of this difference in the energy region
$\omega<0.3$ GeV is problematic. Thus the model dependence of the
asymmetry can experimentally be observed  only at $q^2$ close to
the threshold region, $4m_\pi^2\leq q^2<0.4$ GeV$^2$.

To our view, the photon FSR from the two--pion channel in $e^+ e^-
\to hadrons $ process shows the model dependence only near the
two--pion threshold where the photon energy is large. In the bulk
of energies (0.4 GeV$^2 < q^2 < s$) scalar QED is sufficient to
describe the FSR contribution to $a_\mu^{had,\gamma}$ and the
$C$--odd asymmetry. In that way our results validate recent
calculations \cite{Czyz_03} performed in framework of sQED.

It is plausible that the more complicated many--particle channels
are more sensitive to the chiral dynamics.Another possibility
to test chiral models is the region of the space-like photon
momenta ($Q^2 < 0$). In particular, the virtual Compton scattering
on the pion, $e^- \pi^\pm \to e^- \pi^\pm \gamma$, allows one to
obtain information on the pion polarizabilities (see, e.g.,
\cite{unkmeir_02,Donoghue_93}).

\section*{Acknowledgements}

\hspace{0.5cm} We are  grateful to  S. Eidelman for careful
reading the manuscript and valuable suggestions. We thank J.F. Donoghue for his comments concerning Ref.~\cite{Donoghue_93}.

\begin{appendix}
\section*{Appendix A. General form of FSR tensor}
\label{app:A}
\setcounter{equation}{0}
\def\theequation{A.\arabic{equation}}

\hspace{0.5cm}The amplitude of the reaction $\gamma ^{\ast
}(Q)\rightarrow \gamma (k)+\pi ^{+}(p_{+})+\pi ^{-}(p_{-})$
depends on  three 4-momenta, which can be chosen as $Q,k$, and
$l\equiv p_{+}-p_{-}$. Here $Q=p_{1}+p_{2}$ is the total momentum
of the $e^{+}e^{-}$ pair with $Q^{2}=s=4E^{2}$. For 
on-mass-shell pions  $Q\cdot l=k\cdot l$. In general,
the second-rank Lorentz tensor $M^{\mu \nu }(Q,k,l)$ can be
decomposed through 10 independent tensors
\cite{Tarrach_75,Drechsel_97}:
\begin{eqnarray}
&&M^{\mu \nu }(Q,k,l)=\sum_{i=1}^{10}\Omega _{i}^{\mu \nu
}F_{i}(Q^{2},k^{2},Q\cdot k,k\cdot l),  \label{eq:A1} \\
&&\Omega _{i}^{\mu \nu }=\{g^{\mu \nu },Q^{\mu }Q^{\nu },k^{\mu }k^{\nu
},l^{\mu }l^{\nu },l^{\mu }Q^{\nu },Q^{\mu }l^{\nu },l^{\mu }k^{\nu },k^{\mu
}l^{\nu },Q^{\mu }k^{\nu },k^{\mu }Q^{\nu }\},  \nonumber
\end{eqnarray}
where parity conservation is taken into account. The tensor $M^{\mu \nu
}(Q,k,l)$ obeys the following properties:
\begin{equation}
M^{\mu \nu }(Q,k,l)=M^{\mu \nu }(Q,k,-l)=M^{\nu \mu }(-k,-Q,l).
\label{eq:A2}
\end{equation}
The first equality follows from the charge conjugation symmetry of
the S-matrix element $\langle \gamma (k),\pi ^{+}(p_{+})\pi
^{-}(p_{-})|S|\gamma ^{\ast }(Q)\rangle =\langle \gamma (k),\pi
^{-}(p_{+})\pi ^{+}(p_{-})|S|\gamma ^{\ast }(Q)\rangle $, and the
second one is due to the photon crossing symmetry:
$Q\leftrightarrow -k$ and $\mu \leftrightarrow \nu $. \
Eqs.~(\ref{eq:A2}) impose certain constraints on the scalar
functions $ F_{i}(Q^{2},k^{2},Q\cdot k,k\cdot l)$.

The consideration below follows that of Ref.~\cite{Drechsel_97},
where the virtual Compton scattering $\gamma ^{\ast }(q)+\pi
^{+}(p)\rightarrow \gamma (q^{\prime })+\pi ^{+}(p^{\prime })$
with the space-like initial photon ($ q^2 < 0$) and real final
photon ($q^{\prime2}=0$) has been studied in detail. Some of the
results for the reaction $\gamma ^{\ast }(Q)\rightarrow \gamma
(k)+\pi ^{+}(p_{+})+\pi ^{-}(p_{-})$ can be obtained from the
corresponding results of \cite{Drechsel_97} after the
substitutions: $p^{\mu }\rightarrow -p_{-}^{\mu }$, $p^{\prime \mu
}\rightarrow p_{+}^{\mu }$, $q^{\mu }\rightarrow Q^{\mu }$,
$q^{\prime \mu }\rightarrow k^{\mu }$.

The gauge invariance conditions $Q_{\mu }M^{\mu \nu }(Q,k,l)=0$
and $M^{\mu \nu }(Q,k,l)k_{\nu }=0$ \ lead to the five linear
equations between functions the  $F_{i}$ in Eq.~(\ref{eq:A1}), and in the
general case of two virtual photons one is left with five
scalar functions (see Eqs.~(14) and (15) of \cite{Drechsel_97}).
We are interested in the situation, where the final photon is
real, i.e. $k^{2}=0$ and $k^{\nu }\epsilon _{\nu }^{\prime }=0$
($\epsilon _{\nu }^{\prime }$ is the polarization vector of the
final photon), while the initial photon produced in $e^{+}e^{-}$
annihilation is virtual with $Q^{2}\geq 4m_{\pi}^{2}$. In this
case one obtains
\begin{equation}
M^{\mu \nu }(Q,k,l)=-ie^{2}(\tau _{1}^{\mu \nu }f_{1}+\tau
_{2}^{\mu \nu }f_{2}+\tau _{3}^{\mu \nu }f_{3})\equiv
-ie^{2}M_{F}^{\mu \nu }(Q,k,l), \label{eq:A3}
\end{equation}
with the gauge invariant tensors
\begin{eqnarray}
\tau _{1}^{\mu \nu } &=&k^{\mu }Q^{\nu }-g^{\mu \nu }k\cdot Q,  \nonumber \\
\tau _{2}^{\mu \nu } &=&k\cdot l(l^{\mu }Q^{\nu }-g^{\mu \nu }k\cdot
l)+l^{\nu }(k^{\mu }k \cdot l-l^{\mu }k \cdot Q),  \nonumber \\
\tau _{3}^{\mu \nu } &=&Q^{2}(g^{\mu \nu }k\cdot l-k^{\mu }l^{\nu
})+Q^{\mu }(l^{\nu }k\cdot Q-Q^{\nu }k\cdot l).
\label{eq:A4}
\end{eqnarray}
The scalar functions $f_{i}\equiv f_{i}(Q^{2},k\cdot Q,k\cdot l)$
are either even or odd with respect to the change of sign of the
argument $k\cdot l$:
\begin{eqnarray}
f_{1,2}(Q^{2},k\cdot Q,k\cdot l) &=&+f_{1,2}(Q^{2},k\cdot Q,-k\cdot l),
\nonumber \\
f_{3}(Q^{2},k\cdot Q,k\cdot l) &=&-f_{3}(Q^{2},k\cdot Q,-k\cdot l).
\label{eq:A5}
\end{eqnarray}
The factor $-ie^{2}$ in Eq.~(\ref{eq:A3}) is included for convenience.
It thus follows that the evaluation of the FSR tensor amounts to
the calculation of the scalar functions $f_{i}$ \ ($i=1,2,3$).
\end{appendix}

\begin{appendix}
\section*{Appendix B. Chiral Lagrangian for pseudoscalar, vector and
axial-vector mesons and photon} \label{app:B}

\setcounter{equation}{0}
\def\theequation{B.\arabic{equation}}

\hspace{0.5cm}We choose the $\mathcal{O}(p^{2})$ chiral Lagrangian
derived by Ecker et al. \cite{Ecker_89}, which contains vector
mesons and axial--vector mesons.  It includes interaction of
pseudoscalar, vector and axial--vector mesons, and photon. The
explicit form is given in \cite{Ecker_89}:
\begin{eqnarray}
L &=&\frac{f_{\pi }^{2}}{4}\mathrm{Tr} (D_{\mu
}UD^{\mu }U^{\dagger} +\chi U^{\dagger} +\chi^{\dagger} U)
-\frac{1}{4}F_{\mu \nu} F^{\mu \nu}
\nonumber \\
&&
- \frac{1}{2} \mathrm{Tr}(\nabla^\lambda V_{\lambda \mu}
\nabla_\nu V^{\nu \mu} - \frac{1}{2}m^2_{\rho} V_{\mu \nu} V^{\mu
\nu} ) - \frac{1}{2} \mathrm{Tr}(\nabla^\lambda A_{\lambda \mu}
\nabla_\nu A^{\nu \mu}
- \frac{1}{2}m^2_{a} A_{\mu \nu} A^{\mu \nu} ) \nonumber \\
&&
+\frac{F_{V}}{2\sqrt{2}}\mathrm{Tr}(V_{\mu \nu }f_{+}^{\mu \nu })+\frac{%
iG_{V}}{\sqrt{2}}\mathrm{Tr}(V_{\mu \nu }u^{\mu }u^{\nu })+\frac{F_{A}}{2%
\sqrt{2}}\mathrm{Tr}(A_{\mu \nu }f_{-}^{\mu \nu }),  \label{eq:B1}
\end{eqnarray}
where $U=\exp (i\sqrt{2}\Phi /f_{\pi })$, \ $\Phi $ describes
the $SU(3)$ octet of pseudoscalar mesons, $V_{\mu \nu } \ (A_{\mu \nu
})$ is antisymmetric field describing the $SU(3)$\ octet of
polar-vector (axial-vector) mesons, and $F^{\mu \nu }=\partial
^{\mu }B^{\nu }-\partial ^{\nu }B^{\mu }$ is the EM tensor, where
the photon field is denoted by $B^\mu$. Further,  $ f_{\pi },\
F_{V},G_{V},F_{A}$ are constants, whose numerical values are
specified in Sect.~\ref{sec:results}. For more details on
definitions and notation see Ref.~\cite{Ecker_89}.

For treating the process $\gamma ^{\ast }\rightarrow \pi ^{+}\pi
^{-}\gamma $ at tree level it is sufficient to keep in
Eq.~(\ref{eq:B1}) the terms containing the neutral meson $\rho
^{0} (770)$, and the charged mesons $ a_{1}^{\pm} (1260)$ and
$\pi^{\pm}$, as well as the photon. We obtain
\begin{eqnarray}
L &=&ieB^{\mu }(\pi ^{-}\partial _{\mu }\pi ^{+}-\pi ^{+}\partial _{\mu }\pi
^{-})+e^{2}B_{\mu }B^{\mu }\pi ^{+}\pi ^{-}+e\frac{F_{V}}{2}F^{\mu \nu }\rho
_{\mu \nu }^{0}(1-\frac{\pi ^{+}\pi ^{-}}{f_{\pi }^{2}})  \nonumber \\
&&+i\frac{G_{V}}{f_{\pi }^{2}}\rho _{\mu \nu }^{0}(\partial ^{\mu }\pi
^{+}\partial ^{\nu }\pi ^{-}-\partial ^{\mu }\pi ^{-}\partial ^{\nu }\pi
^{+})+ie\frac{F_{A}}{2f_{\pi }}F^{\mu \nu }(a_{1\mu \nu }^{+}\pi
^{-}-a_{1\mu \nu }^{-}\pi ^{+})  \nonumber \\
&&+e\frac{G_{V}}{f_{\pi }^{2}}\rho _{\mu \nu }^{0}[B^{\mu }(\pi
^{+}\partial ^{\nu }\pi ^{-}+\pi ^{-}\partial ^{\nu }\pi
^{+})-B^{\nu }(\pi ^{+}\partial ^{\mu }\pi ^{-}+\pi ^{-}\partial
^{\mu }\pi ^{+})].  \label{eq:B3}
\end{eqnarray}
Lagrangian (\ref{eq:B3}) leads to the Feynman rules discussed in
Appendix~C. The diagrams contributing to the $ \gamma ^{\ast
}\rightarrow \pi ^{+}\pi ^{-}\gamma $ reaction at tree level
are shown in Fig.~2. For a general case of the two virtual photons
we obtain the FSR tensor $M_{F}^{\mu \nu }$:
\begin{eqnarray}
M_{F}^{\mu \nu } &=&\frac{1}{Q^{2}-2Q\cdot p_{-}}\Gamma ^{\mu
}(Q-p_{-},p_{-})\Gamma ^{\nu }(p_{+},p_{+}+k)+(p_{+}\leftrightarrow
p_{-})-2g^{\mu \nu }  \nonumber \\
&&+\frac{F_{V}^{2}-2F_{V}G_{V}}{f_{\pi }^{2}}\biggl(
\frac{1}{m_{\rho }^{2}-Q^{2}}+\frac{1}{m_{\rho }^{2}-k^{2}}\biggr)
(g^{\mu \nu }k \cdot Q-k^{\mu
}Q^{\nu })  \nonumber \\
&&-\frac{2F_{V}G_{V}}{f_{\pi }^{2}}\left[ \frac{1}{m_{\rho
}^{2}-Q^{2}} (g^{\mu \nu }Q^{2}-Q^{\mu }Q^{\nu })+\frac{1}{m_{\rho
}^{2}-k^{2}}(g^{\mu
\nu }k^{2}-k^{\mu }k^{\nu })\right]   \nonumber \\
&&+\frac{F_{A}^{2}}{f_{\pi }^{2}m_{a}^{2}}\biggl[ (g^{\mu \nu
}k\cdot Q-k^{\mu }Q^{\nu })+\frac{1}{m_{a}^{2}-(Q-p_{-})^{2}} [
(k+p_{+})^{\mu
}(k\cdot Q(k+p_{+})^{\nu }  \nonumber \\
&&-k\cdot (k+p_{+}))-Q\cdot (k+p_{+})(k^{\mu }(k+p_{+})^{\nu
}-k\cdot (k+p_{+})g^{\mu \nu }) ]+(p_{+}\leftrightarrow
p_{-})\biggr] \label{eq:B4}
\end{eqnarray}
where the EM vertex for the off-mass-shell pion (with initial
$p_i$ and final $p_f$ momenta) is
\begin{equation}
\Gamma ^{\mu }(p_{f},p_{i})=(p_{i}+p_{f})^{\mu
}+\frac{F_{V}G_{V}}{f_{\pi }^{2}(m_{\rho
}^{2}-r^{2})}[(p_{i}+p_{f})^{\mu }r^{2}-r^{\mu
}(p_{f}^{2}-p_{i}^{2})],\; \;\;\;\;\;\; r\equiv p_{f}-p_{i}.
\label{eq:B5}
\end{equation}
Note that Lagrangian (\ref{eq:B1}) was applied in
Ref.~\cite{Donoghue_93} in calculation of the Compton tensor for
$\gamma ^{\ast }\pi ^{+}\rightarrow \gamma ^{\ast }\pi ^{+}$. 
\end{appendix}

\begin{appendix}
\section*{Appendix C. Feynman rules for ChPT Lagrangian}
\label{app:C}

\setcounter{equation}{0}
\def\theequation{C.\arabic{equation}}


\hspace{0.5cm}Following \cite{Ecker_89} we describe the vector
(axial-vector) meson by the antisymmetric tensor field that
corresponds to the following form of the propagator
\begin{equation}
i\Delta^{\alpha\beta;\mu\nu}(q) =
\frac{i}{M^2(M^2-q^2)}[g^{\alpha\mu}g^{\beta\nu}(M^2-q^2)+g^{\alpha\mu}q^\beta
q^\nu-g^{\alpha\nu}q^\beta q^\mu-(\alpha\leftrightarrow\beta)] ,
\end{equation}
where $M$ is a mass of the vector (axial-vector) meson.

The vertices corresponding to the ChPT Lagrangian from Appendix~B
are listed in Fig.~7.

\begin{figure}
\begin{center}
\epsfig{file=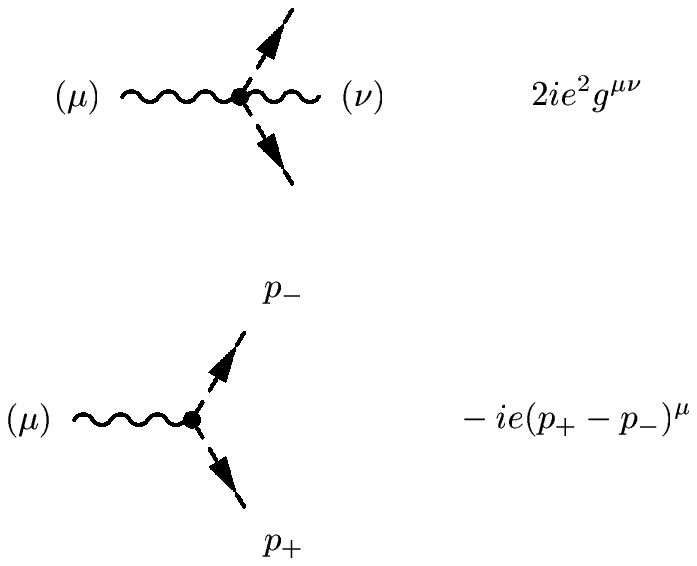,width=5.2cm,height=5.5cm} \label{figApp1}
\hspace{1cm} \epsfig{file=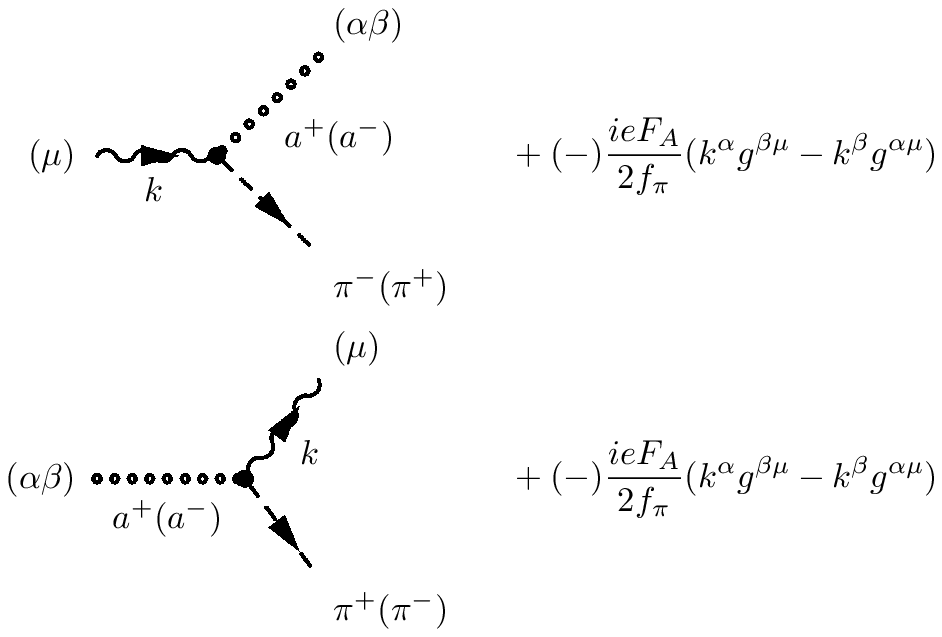,width=6.5cm,height=5.9cm}

\vspace{1.5cm} \epsfig{file=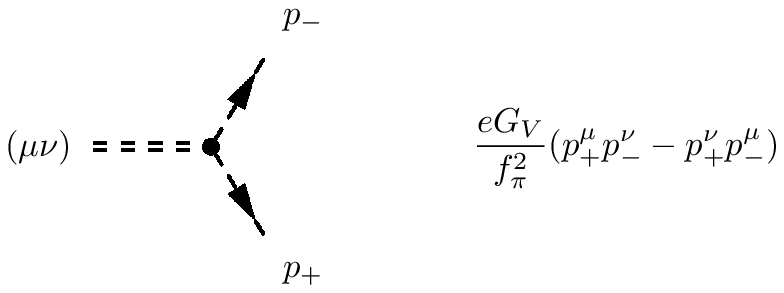,width=5.5cm,height=3.1cm}
\label{figApp2} \hspace{1cm}
\epsfig{file=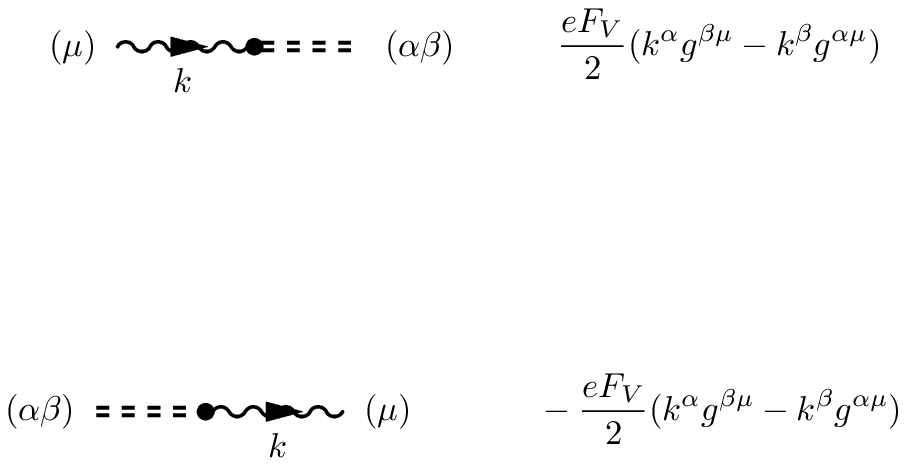,width=6cm,height=3.5cm} \label{figApp4}

\vspace{1cm}\epsfig{file=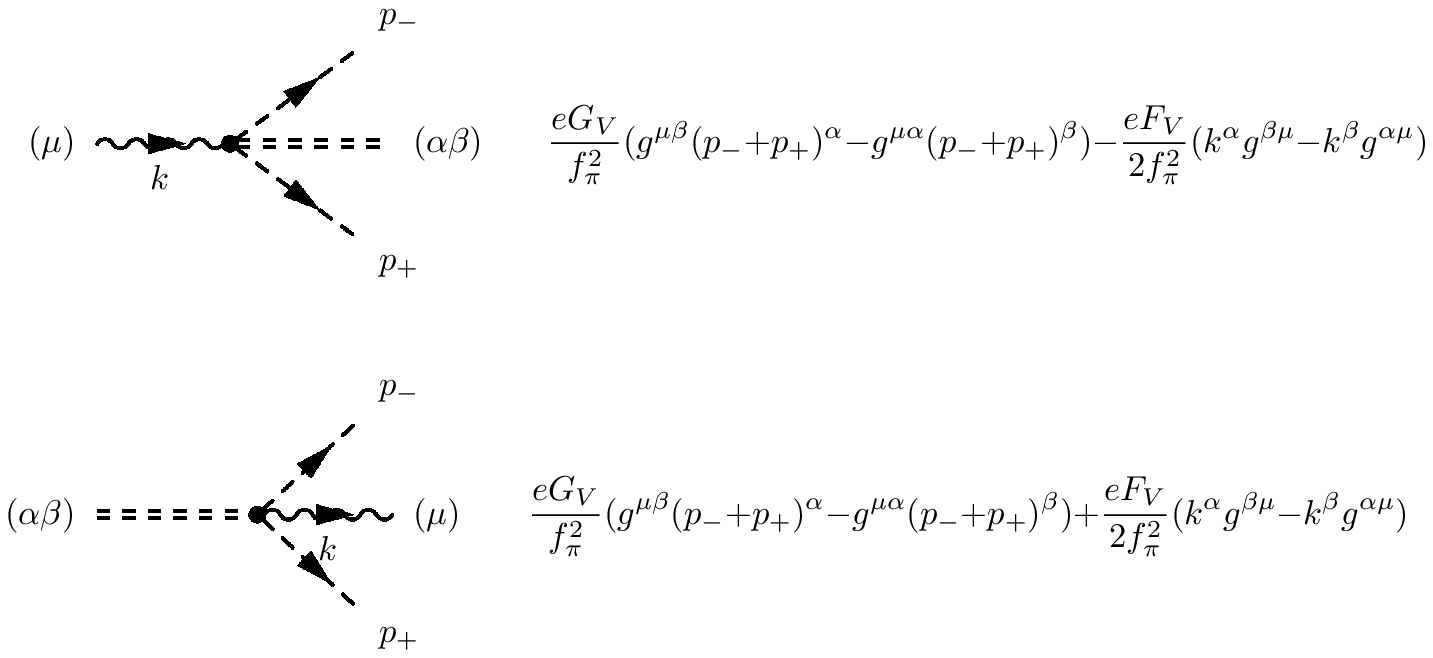,width=9cm,height=6cm}
\label{figApp3}
\end{center}
\caption{}
\end{figure}

\bigskip

\bigskip

\end{appendix}

\begin{appendix}
\section*{Appendix D. Expressions for functions $h_{1,2}$}
\label{app:D}

\setcounter{equation}{0}
\def\theequation{D.\arabic{equation}}

\hspace{0.5cm} The functions $h_1$ and $h_2$ of
Sect.~\ref{subsec:final-state} can be written as
\begin{equation}
h_i=h_i^{sQED}+\Delta h_i,
\end{equation}
where $h_i^{sQED}$ are determined from the following equations
\cite{Baier_65}
\begin{equation}
h_2^{sQED} (k\cdot q)^2=4\pi \xi |F_\pi(Q^2)|^2
[3q^2-(q^2+2m_\pi^2) L_1 ]
\end{equation}
\begin{eqnarray}
2h_1^{sQED}-h_2^{sQED}Q^2=\frac{16\pi\xi|F_\pi(Q^2)|^2}{(k\cdot
Q)^2}
[(k\cdot Q)^2+( \frac{Q^2}{4}-m_\pi^2)(-q^2+%
(q^2-2m_\pi^2)L_1)]
\end{eqnarray}
with $\xi = \Bigl( 1-\displaystyle\frac{\mathstrut 4m_\pi^2}{q^2}
\Bigr)^{1/2}$ and  $L_1=\displaystyle\frac{\mathstrut 1}{ \xi}
\ln\frac{1+\xi}{1-\xi}$.

From Eqs.~(\ref{eq:delta-f1})--(\ref{eq:functions-in-ChPT}) we
obtain the equation for $\Delta h_2$ and $(2\Delta h_1-s h_2)$ in
ChPT
\begin{eqnarray}
&& 2\Delta h_1-s\Delta h_2= 2\pi \xi \Biggl\{C_1+\frac{C_2
L_2}{k\cdot Q}+ \frac{C_3}{[(m_a^2-m_\pi^2)(r-k\cdot Q)
+\frac{4m_\pi^2(k\cdot
Q)^2}{q^2}]}+ 4ReF_\pi(s)  \nonumber \\
&& \times\Biggl[f Q^2-a r + 2m_\pi^2\Bigl(-f
 +a\frac{2(m_a^2-m_\pi^2)(r-k\cdot Q)
+2m_\pi^2 r +k \cdot Q(m_a^2-m_\pi^2)}{%
(m_a^2-m_\pi^2)(r-k\cdot Q)}\Bigr)L_1  \nonumber \\
&& - \Biggl(\frac{4m_\pi^2(m_a^2+m_\pi^2)(k \cdot
Q)^2}{(m_a^2-m_\pi^2) (r-k\cdot Q)} +2m_\pi^2
q^2+(m_a^2-m_\pi^2)(r+2Q^2-k\cdot Q)\Biggr)\frac{a L_2}{k\cdot Q%
}\Biggr]\Biggr\} ,
\end{eqnarray}
where
\begin{eqnarray}
C_1&=&2(k\cdot Q)^2 f^2+a f (2q^2 k\cdot Q+r Q^2)+%
\frac{1}{4}a^2((q^2-2k\cdot Q) q^2+2 r^2) ,
\end{eqnarray}
\begin{eqnarray}
C_2&=& -\frac{a f}{2} \{ (k\cdot Q)^2(Q^2+4k\cdot Q
+4m_\pi^2)+2k\cdot
Q(Q^2+2k\cdot Q)r  \nonumber \\
&& +Q^2r^2\} +\frac{a^2}{8r}\{(k\cdot
Q)^2[8m_\pi^4+2m_\pi^2(Q^2+6k\cdot Q)  \nonumber
\\
&& +k\cdot Q(2Q^2+5k\cdot Q)]+2 k\cdot Q r [-q^2 Q^2+12 k\cdot
Q(m_\pi^2+k\cdot
Q)]  \nonumber \\
&& -2r^2[Q^4+m_\pi^2 q^2+k\cdot Q (Q^2-11k\cdot Q)] -4Q^2 r^3
-3r^4\} ,
\end{eqnarray}
\begin{eqnarray}
C_3&=&\frac{a^2}{4}\{ [8m_\pi^4+2m_\pi^2(-Q^2+6k\cdot Q)+5(k\cdot
Q)^2+Q^4](k\cdot Q)^2
+2r k\cdot Q  \nonumber \\
&&\times [Q^4+2Q^2 k\cdot Q+4(k\cdot Q)^2+4m_\pi^2 (k\cdot Q)]
+r^2[Q^4+8Q^2 k\cdot Q  \nonumber \\
&& + 2(k\cdot Q)^2+2m_\pi^2 q^2] +4Q^2r^3+r^4 \},
\end{eqnarray}

$$\Delta h_2 (k \cdot Q)^2=\frac{\pi a (m_a^2-m_\pi^2)\xi
}{2}\Biggl\{q^2(\frac{3a(m_a^2-m_\pi^2)}{2}+8\mathrm{Re}
F_\pi(s))+\frac{16 (k \cdot Q)^2 m_\pi^2 Re F_\pi(s)}{(m_a^2-m_%
\pi^2)(r-k\cdot Q)} L_1-$$

\[
\Biggl[\frac{3q^2(m_a^2-m_\pi^2)(r-k\cdot Q)+2(k\cdot
Q)^2(q^2+2m_\pi^2)}{4 r k\cdot Q}a(m_a^2-m_\pi^2) +
\]
\begin{equation}
8ReF_\pi(s)\frac{r (q^2(m_a^2-m_\pi^2)(r-k\cdot Q)
+4(k\cdot Q)^2m_\pi^2)}{2 k\cdot Q (r-k\cdot Q) (m_a^2-m_\pi^2)}\Biggr]%
L_2\Biggr\} ,
\end{equation}

\[
L_2=\frac{1}{\xi} \ln%
\frac{r+\xi k \cdot Q}{r-\xi k \cdot Q} , \; \; \; \; \; \; r=
m_a^2-m_\pi^2-k\cdot Q ,\]

\[
a=\frac{F_A^2}{m_a^2 f_\pi^2} , \; \; \; \; \; f=\frac{%
F_V^2-2F_V G_V}{f_\pi^2}
\Biggl(\frac{1}{m_\rho^2}+\frac{1}{m_\rho^2-s-
im_\rho\Gamma_\rho}%
\Biggr) .
\]

\end{appendix}

\begin{appendix}
\section*{Appendix E. Kinematics of the 3-particle final state}
\label{app:E}

\setcounter{equation}{0}
\def\theequation{E.\arabic{equation}}

Solving energy-momentum conservation for the pion energy $E_+$ in
Eq.~(\ref{eq:E+}) requires some care. Firstly, notice that the
energy of the photon at fixed CM energy $\sqrt{s}=2E$ varies
within the limits $0 \leq \omega \leq \omega_{max}=E - m_\pi^2 /
E$. Secondly, by requiring positiveness of the expression under
the square root in Eq.~(\ref{eq:E+}) at arbitrary $\cos
\theta_{\gamma+}$, we get the conditions
\begin{equation}
\omega \leq \omega_- = \frac{2E (E -m_\pi )}{2E -m_\pi } ,
\;\;\;\;\;\;\; \omega \geq \omega_{+} = \frac{2E (E+m_\pi )
}{2E+m_\pi} .
\end{equation}
Clearly $\omega_+ \geq \omega_{max}$ and $\omega_-  \leq
\omega_{max}$. Thus the restriction for the photon energy is $0
\leq \omega \leq \omega_-$. The corresponding invariant mass of
the $\pi^+ \pi^-$ pair is $ 4E^2 \geq q^2 \geq 4 E^2 m_\pi /(2E
-m_\pi) \approx 2E m_\pi$.

The requirement $0 \leq \omega \leq \omega_-$ coincides with the
condition that the energy conservation low in
Eq.~(\ref{eq:energy-conservation}) leads to one solution for
$E_+$, namely the one in Eq.~(\ref{eq:E+}). In the other case, if
$ \omega_- \leq \omega \leq \omega_{max}$, the situation becomes
more complicated: there appear two solutions $E_{+}^{(-)}$ and
$E_{+}^{(+)}$, which differ by the sign in front of the square
root in Eq.~(\ref{eq:E+}). Correspondingly one has to sum over two
terms in Eq.~(\ref{eq:energy-conservation}) corresponding to these
solutions. Besides, the angle $\theta_{\gamma+}$ in this case is
limited by the value
\begin{equation}
(\sin^2 \theta_{\gamma+})_{max} = \frac{4E (E-\omega)
[E(E-\omega)-m_\pi^2]}{m_\pi^2 \omega^2 } .
\end{equation}
For these values of photon energies each angle $\theta_{\gamma+}$
in the Lab frame (CM frame for colliding $e^+ e^-$ beams)
corresponds to the two different angles between momenta of $\pi^+$
and $\gamma$ in the $\pi^+ \pi^-$ CM frame.
 Here we refer to monograph \cite{Byckling_73} (ch. III),
where these aspects of kinematics are considered in detail.

\end{appendix}

\begin{appendix}
\section*{Appendix F. Contribution to FSR from $\rho^\pm \to \pi^\pm \gamma$ decays }
\label{app:F}

\setcounter{equation}{0}
\def\theequation{F.\arabic{equation}}

The diagrams with intermediate charged $\rho$-meson can be
obtained from the 3rd row of diagrams in Fig.~2, if
$a_{1}^\pm$--meson is replaced by $\rho^\pm$--meson.
%
We choose the chiral 
Lagrangian, describing the odd-intrinsic-parity sector, in the
form \cite{Pich_90}
\begin{equation}
L^{(V)} =H_V\epsilon_{\mu \nu \alpha \beta}
\mathrm{Tr}(V^\mu\{u^\nu,f_+^{\alpha\beta}\})\simeq
-\frac{2\sqrt{2}eH_V}{3f_\pi} \epsilon_{\mu \nu \alpha \beta}
F^{\alpha\beta} \vec{\rho}^\mu \partial^\nu{\vec\pi}
\label{eq:app-rho-pi-gamma}
\end{equation}
in the vector formulation for the $\rho$--meson field, where
$\epsilon_{\mu \nu \alpha \beta}$ is the totally antisymmetric
Levi-Civita tensor.

The constant $H_V$ can be determined from the $\rho^\pm \to
\pi^\pm \gamma$ decay width $\Gamma(\rho^\pm \to \pi^\pm
\gamma)=68.7$ keV \cite{PDG}. From Eq.~(\ref{eq:app-rho-pi-gamma})
one finds
\begin{equation}
\Gamma(\rho^\pm \to \pi^\pm \gamma) = \frac{4 \alpha m_\rho^3
H_V^2 }{27 f_\pi^2 }(1- \frac{m_\pi^2}{m_\rho^2})^3
\label{eq:app-width-rho}
\end{equation}
and thus $H_V=0.0363$.

The corresponding contribution to the invariant functions of
Sect.~\ref{sec:final-state-ChPT} takes the form
\begin{eqnarray}
\Delta f_{1}^{\rho^\pm} &=&\frac{8 H_{V}^{2}}{9 f_{\pi }^{2}}
\biggl[ ({k\cdot Q +l^2}) \biggl( \frac{1}{C(l)}+ \frac{1}{C(-l)}
\biggr) + 2 k \cdot l \biggl( \frac{1}{C(l)} - \frac{1}{C(-l)}
\biggr) \biggr] ,
\label{eq:app-delta-f1} \\
\Delta f_{2}^{\rho^\pm} &=&-\frac{8 H_{V}^{2}}{9 f_{\pi }^{2}}
\biggl( \frac{1}{C(l)}+ \frac{1}{C(-l)} \biggr) ,
\label{eq:app-delta-f2} \\
\Delta f_{3}^{\rho^\pm} &=&\frac{8 H_{V}^{2}}{9 f_{\pi }^{2}}
\biggl( \frac{1}{C(l)} - \frac{1}{C(-l)} \biggr) ,
\label{eq:app-functions-in-ChPT}
\end{eqnarray}
where \ $C(\pm l) =m_{\rho}^{2}- (k+ p_{\pm})^2 -im_{\rho
}\Gamma_\rho ((k+ p_{\pm})^2)$ \ with \ $(k+ p_{\pm})^2 =
(Q^{2}+l^{2}+2k\cdot Q \pm 4k\cdot l)/4$.

If we choose the antisymmetric--tensor field formulation for the
$\rho$--meson as was done in the rest of this paper, then the
Lagrangian, which is equivalent to (\ref{eq:app-rho-pi-gamma}) on the
mass shell, reads
\begin{equation}
L^{(T)} =\frac{\sqrt{2}e H_V m_\rho}{3 f_\pi} \epsilon_{\mu \nu
\alpha \beta} F^{\alpha\beta} \vec{\rho}^{\mu \nu } \vec{\pi}   .
\label{eq:app-1rho-pi-gamma}
\end{equation}
For this Lagrangian the functions $\Delta f_{2,3}^{\rho^\pm}$ are
the same as in (\ref{eq:app-delta-f2}) and
(\ref{eq:app-functions-in-ChPT}), while $\Delta f_{1}^{\rho^\pm}$
differs from (\ref{eq:app-delta-f1}) by an additional term,
\begin{equation}
(\Delta f_{1}^{\rho^\pm })^\prime = \Delta f_{1}^{\rho^\pm} +
\frac{64 H_V^2}{9 f_\pi^2} . \label{eq:app-delta-f1-tensor}
\end{equation}

\begin{figure}
\begin{center}
\epsfig{file=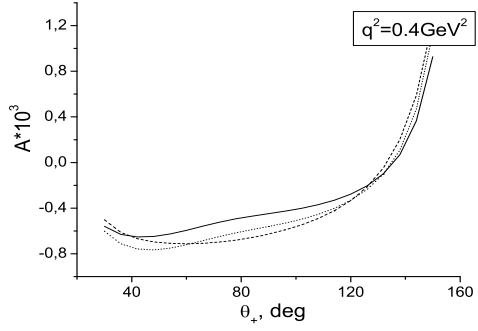,width=7cm,height=5cm} \label{ro}
\vspace{-.5cm} \caption{Charge asymmetry as a function of pion
polar angle for $s=1$ GeV$^2$. The solid (dotted) line corresponds
to the tensor (vector) formulation for $\rho$--meson, the dashed
line  -- calculation without $\rho\to\pi\gamma$ contribution.}
\end{center}
\end{figure}

According to our calculations at invariant masses from the
two--pion threshold to $q^2 \approx 0.4$ GeV$^2$ the
$\rho\to\pi\gamma$ contribution to the charge asymmetry may be of
the same order as the $a_1 \to \pi\gamma $ contribution, if the
tensor formulation for the $\rho$--meson field is applied (see
Fig.~8). For the higher values of $q^2$ the considered mechanism
is suppressed with respect to other contributions.

Regarding the seeming difference between vector and tensor
formulations, we should note that, as argued in
\cite{Pich_90,Bijnens_96}, the effective Lagrangians in the two
formulations would become equivalent 
if the Lagrangian in the tensor formulation included an additional
local term. Apparently the contribution from this local term to
the functions $\Delta f_i^{\rho^\pm}$ would cancel the term $64
H_V^2/(9 f_\pi^2)$ in (\ref{eq:app-delta-f1-tensor}) making the
charge asymmetry independent of the formulation for the
$\rho$--meson field. Therefore we can conclude that the
contribution of the $\gamma^* \to \rho^\pm \pi^\mp \to \pi^+ \pi^-
\gamma$ process to the asymmetry is very small at all two--pion
invariant masses.
\end{appendix}


\end{document}